\documentclass[journal,draftcls,onecolumn,12pt,twoside]{IEEEtran}
\usepackage{graphicx,multirow}
\usepackage{epsfig,amsfonts}
\usepackage{psfrag}
\usepackage{dsfont}
\usepackage{cancel}
\usepackage{subcaption}
\usepackage{nicefrac}
\usepackage{manfnt}
\usepackage{cite}
\usepackage{amssymb,eucal}
\usepackage{exscale}
\usepackage{verbatim}
\usepackage{amstext}
\usepackage{latexsym}
\usepackage{ifthen}
\usepackage{fancybox}
\usepackage{gensymb}
\usepackage{amsthm}
\usepackage{amsmath}
\usepackage{color}
\usepackage{bbm}
\usepackage{epstopdf}
\usepackage[table,xcdraw]{xcolor}

\newtheorem{lemma}{\textbf{\text{Lemma}}}

\newtheorem{approximation}{Approximation}
\newtheorem{remark}{Remark}
\usepackage{collcell, datatool}
\usepackage{pifont}

\definecolor{lightgray}{gray}{0.9}
\usepackage{algorithm}
\usepackage[algo2e]{algorithm2e}
\usepackage{balance}
\usepackage{ifmtarg}
\usepackage[labelsep=none]{caption}
\usepackage{footnote}
\usepackage{stfloats}

\usepackage{mathrsfs}
\usepackage{tikz}
\usepackage{pgfplotstable}
\usepackage{rotate}

\definecolor{rubblue}{cmyk}{1,0.5,0,0.6}
\definecolor{rubgreen}{cmyk}{0.5,0,1,0}
\definecolor{rubgray}{cmyk}{0.03,0.03,0.03,0.1}

\usepgfplotslibrary{units}

\usetikzlibrary{%
patterns,%
calc,%
fit,%
arrows,%
plotmarks,%
shadows,%
chains,%
shapes%
}

\tikzset{>=latex'} 
\tikzstyle{every picture}+=[remember picture] 
\pgfdeclarelayer{background}
\pgfdeclarelayer{foreground}
\pgfsetlayers{background,main,foreground}

\tikzstyle{blueblock}=[draw=rubblue, rectangle, thick, drop shadow, minimum width=20mm, minimum height=8mm,fill=rubblue!20, text width=20mm, text centered]
\tikzstyle{bluebox}=[draw=rubblue, rectangle, thick, drop shadow, minimum width=8mm, minimum height=8mm,fill=rubblue!20, text width=8mm, text centered]
\tikzstyle{greenblock}=[draw=rubgreen, rectangle, thick, drop shadow, minimum width=20mm, minimum height=8mm,fill=rubgreen!20, text width=20mm, text centered]
\tikzstyle{dot} = [draw, circle, minimum size=0.2pt,scale=0.3,fill=black,black]
\tikzstyle{smalldot} = [draw, circle, minimum size=0.1pt,scale=0.2,fill=black,black]
\tikzstyle{reddot}  =[draw,circle,minimum size=0.2pt,scale=0.8,fill=red,thin]
\tikzstyle{greendot}  =[draw,circle,minimum size=0.2pt,scale=0.8,fill=Green,thin]
\tikzstyle{bluedot}  =[draw,circle,minimum size=0.2pt,scale=0.8,fill=blue,thin]
\tikzstyle{whitedot}=[draw,circle,minimum size=0.2pt,scale=0.8,fill=white,thin]
\tikzstyle{blackdot} = [draw, circle, minimum size=0.2pt,scale=0.7,fill=black,black]
\tikzstyle{sum} = [drop shadow, draw=rubblue, thick, fill=rubblue!20, circle]
\tikzstyle{relay} = [blueblock, minimum width=5mm, minimum height=20mm, text width=5mm, rounded corners=2pt]
\tikzstyle{relay2} = [blueblock, minimum width=5mm, minimum height=15mm, text width=5mm, rounded corners=2pt]
\tikzstyle{relay3} = [blueblock, minimum width=5mm, minimum height=25mm, text width=5mm, rounded corners=2pt]
\tikzstyle{relay4} = [blueblock, minimum width=5mm, minimum height=10mm, text width=5mm, rounded corners=2pt]
\tikzstyle{relay5} = [blueblock, minimum width=5mm, minimum height=50mm, text width=5mm, rounded corners=2pt]
\tikzstyle{relay6} = [blueblock, minimum width=5mm, minimum height=5mm, text width=5mm, rounded corners=2pt]
\tikzstyle{circgreen} = [draw, circle, inner sep=2pt, fill=rubgreen, drop shadow, thick]
\tikzstyle{circwhite} = [draw, circle, inner sep=2pt, fill=white, drop shadow, thick]
\tikzstyle{circdashed} = [draw, dashed, circle, inner sep=2pt, fill=rubgray, drop shadow, thick]
\tikzstyle{vertbox} = [rectangle, draw=rubblue, thick, rotate=90, text centered, minimum width=16.5mm, minimum height=8mm, text width=16.5mm, inner sep=0pt, fill=rubblue!20, drop shadow]
\tikzstyle{vertboxb} = [rectangle, draw=rubblue, thick, rotate=90, text centered, minimum width=16.5mm, minimum height=8mm, text width=16.5mm, fill=rubblue!20, drop shadow]
\tikzstyle{vertboxshort} = [rectangle, draw=rubblue, thick, rotate=90, text centered, minimum width=10mm, minimum height=8mm, text width=10mm, inner sep=0pt, fill=rubblue!20, drop shadow]
\tikzstyle{smalldotgreen} = [draw=rubgreen, circle, minimum size=0.2pt,scale=0.8,fill=rubgreen!20]
\tikzstyle{antenna} = [regular polygon, regular polygon sides=3, draw, shape border rotate=180, minimum size=0.2pt, scale=0.3]

\tikzstyle{poly} = [regular polygon, regular polygon sides=6, shape aspect=0.5, minimum width=1.5cm, minimum height=0.35cm, draw, dashed]

\definecolor{cff9e00}{RGB}{255,158,0}
\definecolor{c4fff00}{RGB}{79,255,0}
\definecolor{cff0012}{RGB}{255,0,18}
\definecolor{c00c5ff}{RGB}{0,197,255}
\definecolor{c046f00}{RGB}{4,111,0}
\definecolor{c004b9d}{RGB}{0,75,157}

\newlength{\mylen}
\usetikzlibrary{petri}
\usetikzlibrary{shapes}
\usetikzlibrary{positioning,arrows}
\usepackage[utf8x]{inputenc}
\usepackage{scalefnt}
\usetikzlibrary{calc,decorations.markings}

\usepackage[outline]{contour}
\contourlength{1.2pt}

\newcommand{\RN}[1]{%
	\textup{\uppercase\expandafter{\romannumeral#1}}%
}

\usepackage{xparse}
\NewDocumentCommand{\ceil}{s O{} m}{%
	\IfBooleanTF{#1} 
	{\left\lceil#3\right\rceil} 
	{#2\lceil#3#2\rceil} 
}
\NewDocumentCommand{\floor}{s O{} m}{%
	\IfBooleanTF{#1} 
	{\left\lfloor#3\right\rfloor} 
	{#2\lfloor#3#2\rfloor} 
}

\usepackage{amsmath}
\usepackage{algorithm}
\usepackage[noend]{algpseudocode}

\makeatletter
\def\BState{\State\hskip-\ALG@thistlm}
\makeatother

\makeatletter
\newif\if@borderstar
\def\bordermatrix{\@ifnextchar*{%
		\@borderstartrue\@bordermatrix@i}{\@borderstarfalse\@bordermatrix@i*}%
}
\def\@bordermatrix@i*{\@ifnextchar[{\@bordermatrix@ii}{\@bordermatrix@ii[()]}}
\def\@bordermatrix@ii[#1]#2{%
\begingroup
\m@th\@tempdima8.75\p@\setbox\z@\vbox{%
	\def\cr{\crcr\noalign{\kern 2\p@\global\let\cr\endline }}%
	\ialign {$##$\hfil\kern 2\p@\kern\@tempdima & \thinspace %
		\hfil $##$\hfil && \quad\hfil $##$\hfil\crcr\omit\strut %
		\hfil\crcr\noalign{\kern -\baselineskip}#2\crcr\omit %
		\strut\cr}}%
\setbox\tw@\vbox{\unvcopy\z@\global\setbox\@ne\lastbox}%
\setbox\tw@\hbox{\unhbox\@ne\unskip\global\setbox\@ne\lastbox}%
\setbox\tw@\hbox{%
	$\kern\wd\@ne\kern -\@tempdima\left\@firstoftwo#1%
	\if@borderstar\kern2pt\else\kern -\wd\@ne\fi%
	\global\setbox\@ne\vbox{\box\@ne\if@borderstar\else\kern 2\p@\fi}%
	\vcenter{\if@borderstar\else\kern -\ht\@ne\fi%
		\unvbox\z@\kern-\if@borderstar2\fi\baselineskip}%
	\if@borderstar\kern-2\@tempdima\kern2\p@\else\,\fi\right\@secondoftwo#1 $%
}\null \;\vbox{\kern\ht\@ne\box\tw@}%
\endgroup
}

\begin{document}


\title{Grant-Free {Opportunistic} Uplink Transmission in Wireless-powered IoT: A Spatio-temporal   Model
}
\author{
	\IEEEauthorblockN{\large  Mohammad Gharbieh, Hesham ElSawy, {\em Senior Member, IEEE}, Mustafa Emara, {\em Student Member, IEEE}, Hong-Chuan Yang, {\em Senior Member, IEEE} , and Mohamed-Slim Alouini, {\em Fellow, IEEE}\\
		\thanks{The paper is accepted for publication in the IEEE Transactions on Communications on November 11, 2020.\;}
		\thanks{ M. Gharbieh and  H.-C. Yang are with the Department of Electrical and Computer Engineering, University of Victoria, Victoria, BC V8P 5C2, Canada; e-mails: \{mohammadgharbieh, hy\}@uvic.ca. \;}
		\thanks{H. ElSawy is with the Electrical Engineering Department, King Fahd University of Petroleum and Minerals (KFUPM), Dhahran 31261, Saudi Arabia; e-mail: hesham.elsawy@kfupm.edu.sa.\;}	
		\thanks{M. Emara is with the Germany standards R\&D team, Next Generation and
		Standards, Intel Deutschland GmbH and the Institute of Communications,
		Hamburg University of Technology, Hamburg, 21073 Germany (e-mail:
		mustafa.emara@intel.com)\;}
		\thanks{ M.-S. Alouini is with the Computer, Electrical, and Mathematical Sciences and Engineering (CEMSE) Division, King Abdullah University of Science and Technology (KAUST), Thuwal 23955, Saudi Arabia; e-mail: slim.alouini@kaust.edu.sa.  \;}	
}}
\maketitle
\thispagestyle{plain}
\pagestyle{plain}

\begin{abstract}
Ambient radio frequency (RF) energy harvesting is widely promoted as an enabler for wireless-power Internet of Things (IoT) networks. This paper jointly characterizes energy harvesting and packet transmissions in grant-free opportunistic uplink IoT networks energized via harvesting downlink energy. To do that, a joint queuing theory and stochastic geometry  model is utilized to develop a spatio-temporal analytical model. Particularly, the harvested energy and packet transmission success probability are characterized using tools from stochastic geometry. {Moreover, each device is modeled using a two-dimensional discrete-time Markov chain (DTMC). Such two dimensions are utilized to jointly track the scavenged/depleted energy to/from the batteries along with the arrival/departure of packets to/from devices buffers over time.  Consequently, the adopted queuing model represents the devices as spatially interacting queues. To that end, the network performance is assessed in light of the packet throughput, the average delay, and the average buffer size.  The effect of base stations (BSs) densification is discussed and several design insights are provided. The results show that the parameters for uplink power control and opportunistic channel access should be jointly optimized to maximize  average network packet throughput, and hence, minimize delay.}
\begin{IEEEkeywords}
IoT networks, grant-free access, opportunistic transmission, energy harvesting, spatio-temporal model, stochastic geometry, 2D-DTMC.
\end{IEEEkeywords}
\end{abstract}

\section{Introduction}

The Internet of Things (IoT) extends connectivity to a wide variety of devices, such as sensors, actuators, smart objects, machines, and vehicles to enable a plethora of applications such as crowd and venue management, ubiquitous health-care, public safety, environmental monitoring, and industrial automation~\cite{7544470}. As such, the IoT is expected to be massive in terms of the spatial span and the number of connected devices \cite{first.mile.KAUST}. {Materializing the IoT paradigm involves several acute challenges in terms of network capacity and administration. For instance, monitoring the battery levels of the devices and replacing/recharging depleted batteries involve overwhelming administrative overhead. Furthermore, traffic generated by massive numbers of sporadically active devices cannot be efficiently served via conventional scheduling or random access schemes~\cite{first.mile.KAUST, massive_Ekram, IOT2,6824261, 8491334}. For instance, it is shown that conventional scheduling schemes involve unnecessary overhead (i.e., scheduling request, scheduling feedback, resource allocation) given the occasional data generation per device~\cite{8506623}. Hence, many IoT oriented low-power-wide-area (LPWA) networks (e.g., Sigfox and LoRa) use a one-stage grant-free uplink (GF-UL) data transmission scheme (i.e., random access)~\cite{8883217, Goodbye}. However, the massive numbers of devices lead to overwhelming contention and interference on the common spectrum resources~\cite{Gharbieh_tcom}. Hence, energy efficiency, self-sustainability, and scalable medium access control schemes are deemed necessary to support the surging IoT applications.  }


{Tackling the energy shortage and improving the scalability of medium access control in large scale IoT networks are the research focus of several studies in the literature. In this context, stochastic geometry is indispensable to account for the intrinsic aggregate network interference that naturally evolves as the performance limiting parameter of large-scale and massive networks~\cite{hesham_tutorial}. For instance, the authors in \cite{8214275} utilize stochastic geometry to study and design IoT networks with deep-sleep mode and on-demand wake-up for extreme power saving. However, the solution in \cite{8214275} still relies  on the devices' internal batteries that will eventually get depleted. To alleviate the necessity of battery replacement, ambient radio frequency (RF) energy harvesting is a plausible solution to provide sustainable energy sources for the devices. In this context, the authors in \cite{8536464} investigate the ability of the IoT devices to simultaneously charge their batteries from the aggregate downlink energy and correctly decode the information on the intended downlink signal.  The performance of wireless networks that utilize RF energy harvesting has also be studied in \cite{7110493,7726030, 7880719, 7509608, 7820568, 7056528}. However, the works in \cite{8536464, 7110493,7726030, 7880719, 7509608, 7820568, 7056528} only account for the randomness in the energy harvesting process and overlook  the randomness in data traffic generation. Hence, none of the works in  \cite{8214275, 8536464, 7110493,7726030, 7880719, 7509608, 7820568, 7056528} account for the sporadic IoT traffic.}

{To account for the randomness in packets generation/service in large-scale networks, recent efforts have developed spatio-temporal models that are based on an integrated stochastic geometry and queueing analysis. Particularly, the queueing theory is used to track the temporal evolution of packets arrivals/departures to/from the devices' buffers. Meanwhile, stochastic geometry is used to capture the impact of mutual interference on successful packet departures. The recently developed spatio-temporal models are gaining popularity to characterize latency, scalability, and stability of large scale networks~\cite{8688635, Gharbieh_tcom, 8506623, 7486114, 8439089, 8543568, 8710261, 8371220, 8835898}. For instance, the scalability and stability of random access in IoT networks are characterized in~\cite{8688635, Gharbieh_tcom, 7486114, 8439089, 8543568, ElSawy2020}. The authors in~\cite{8506623} assess and compare scheduled and grant-free access in uplink IoT networks. The delay of different downlink scheduling schemes is assessed and compared in \cite{8710261, 8371220}. Prioritized data service in large-scale IoT networks is characterized in \cite{8835898, Emara2020}. The age of information in IoT networks with different uplink traffic patterns is assessed in \cite{9042825}. However, the work in \cite{8688635, Gharbieh_tcom, 8506623, 7486114, 8439089, 8543568, 8710261, 8371220, 8835898, 9042825} assume that all devices have perpetual energy sources. Hence, none of the spatio-temporal models in \cite{8688635, Gharbieh_tcom, 8506623, 7486114, 8439089, 8543568, 8710261, 8371220, 8835898, 9042825}  capture the interplay between the energy harvesting and traffic generation/services  processes. }

{To the best of the authors' knowledge, the only exception that accounts for the interplay between energy harvesting and traffic generation/departure is \cite{Fatma_EH}. In particular, the authors in \cite{Fatma_EH} study the self-sustainability of device-to-deceive (D2D) communications that are powered by recycling downlink cellular energy. However, the scope of  \cite{Fatma_EH} is limited to ad-hoc networks with constant transmission powers, fixed link distances, and 1-persistent Aloha protocol. Hence, the analysis and design of self-sustainable uplink IoT networks is still an open research problem. }

\subsection{Scope \& Contributions}
{This paper considers GF-UL IoT networks powered by harvesting the aggregate downlink cellular networks energy. The GF-UL data transmission is adopted due to its lower packet delay and faster response to empty the devices' buffers when compared to scheduled uplink transmissions~\cite{8506623}. For effective spectrum access and efficient usage of the harvested energy, power control and opportunistic spectrum access are utilized for packets transmission. The proposed opportunistic scheme is bench-marked  by its equivalent Aloha scheme. Path-loss inversion power control is utilized to ensure unified and acceptable received power levels at the serving BSs. It is shown in \cite{Meta_Elsawy, Meta_Haenggi_Control} that full  path-loss inversion suppresses the performance discrepancies among the existing devices. Moreover, it is argued in \cite{9006936} that the power control in some LPWA IoT networks boils down to the full  path-loss inversion power control. Opportunistic spectrum access avoids uplink transmissions during deep channel fades to alleviate wasting the harvested energy on transmissions that are more likely to fail. Opportunistic spectrum access also relieves aggregate network interference by deferring unnecessary packet transmissions, which are more likely to be retransmitted due to the high failure probability during deep fades. } { To study and design self-sustainable uplink opportunistic channel access, this paper develops a spatio-temporal model to account for traffic generation, energy harvesting, and mutual interference that governs the successful transmissions of the data packets.\footnote{This work is presented in part in \cite{Moh_EH}.}In the proposed framework, stochastic geometry characterizes the harvested energy as well as the spatial intra and inter-cell interference. Meanwhile, queuing theory jointly tracks the states of the data and energy queues. In such spatio-temporal setup, the IoT network is abstracted to spatially interacting queues with transmissions tokens. That is, transmission attempts are only allowed (i.e., tokens are generated) if and only if there is enough energy in the battery and the intended channel gain is above a certain threshold. The success/failure of the transmission attempts is governed by the mutual interference among the simultaneously active devices. The main contributions of this paper are as follows:}
	

 \begin{itemize}

   \item {We propose an opportunistic spectrum access to improve the utilization of the harvested energy and relief aggregate network interference. The proposed opportunistic scheme is bench-marked  by its equivalent Aloha scheme.  }
   
   
      \item {We develop a spatio-temporal model that accounts for packet generation, opportunistic spectrum access, Signal-to-Interference-plus-Noise-Ratio (SINR)-governed packet departures, and location-dependent energy harvesting and power control.}      
      
   \item {The developed mathematical model accounts for the conflicting intra-cell GF-UL transmissions from different devices served by the same BSs.} 
	\item To account for the location impact on the energy harvesting process, a novel equiprobable location-dependent EH-classes categorization process is presented.
	
	\item {We assess the {\em packet throughput}, defined as the average successful transmitted packets per time slot.  We show sensitivity of packet throughput to the power control parameter and a channel gain threshold for uplink transmission.}
\end{itemize}

To the best knowledge of the authors, this is the first paper to develop a mathematical model that accounts for the temporal traffic generation, mutual interference between the devices, and the energy harvesting in large-scale IoT uplink  networks.
\vspace{-8mm}
\subsection{Notation \& Organization}
Along the paper, the math italic font is used  for scalars, e.g., $x$. We denote vectors lowercase math bold font, e.g., $\mathbf{x}$, while matrices  are denoted by uppercase  math bold font, e.g.,  $\mathbf{X}$. The identity matrix of size $m \times m$ is denoted as $\mathbf{I}_m$ and the zeros matrix is denoted as $\mathbf{0}$.  The calligraphic font, e.g.,  $\mathcal{X}$ is used to represent a random variable (RV). Moreover, $\mathbb{E}_ \mathcal{X}\{\cdot\}$, $f_ \mathcal{X}\{\cdot\}$, $F_{\mathcal{X}}\left(\cdot \right) $,    and $\mathscr{L}_\mathcal{X}\left(\cdot \right)$ denote, respectively, the expectation, the probability density function (PDF), the cumulative distribution function (CDF), and the Laplace Transform (LT) of the PDF of the random variable $\mathcal{X}$. We use $\mathbb{P} \{\cdot\}  $ to denote the probability. $\Gamma(\cdot)$ indicates the Gamma function, $\gamma(\cdot,\cdot)$ is the lower incomplete Gamma function, ${}_2 F_1(\cdot)$ is the Gaussian hypergeometric function, and $\ceil \cdot$ denotes the ceiling function. $\nu$ denotes the Euler-Mascheroni constant, $H_{(\cdot)}$ is the harmonic number, and red$\psi^{(\cdot)}(\cdot)$ is the polygamma function. The imaginary unit is denoted by $j=\sqrt{-1}$ and imaginary component of a complex number is denoted as $\text{Im}\{\cdot\}$. Finally, $(\cdot)_{[i]}$ denotes the value at the $i^{th}$ iteration. 

The remainder of the paper is organized as follows: Section \ref{sec:System_Model} presents the system model and the assumptions. Section \ref{sic:Performance_Analysis} conducts the stochastic geometry and queueing theory analysis. Section \ref{Results} shows numerical and simulations results. Finally, the main results are summarized and the paper is concluded in Section \ref{sec:Conclusions}.

\section{System Modeling \& Assumptions}\label{sec:System_Model}

\begin{figure}[t!]
	\centering
	\includegraphics[width=4 in ]{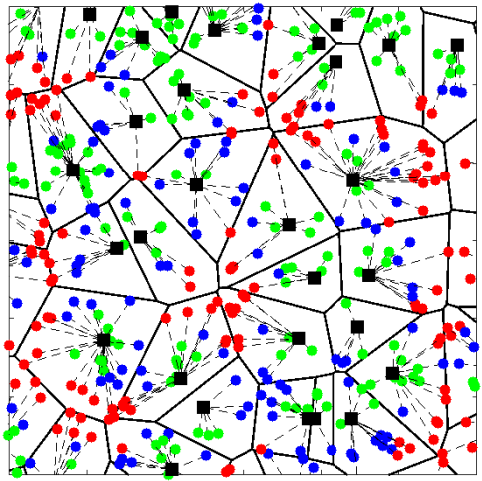}
	\caption{$\;\;$A network realization with the black squares denote the {\rm BS}s and the devices are denoted by the red, green, and blue dots based on their distances to the serving BS.  The solid lines denote the BS's Voronoi cell while the dashed lines denote the device's association to its BS.}\label{Network}
\end{figure}

\subsection{Spatial \& Physical Layer Parameters}

A single-tier of BSs is considered, where the BSs are spatially distributed according to homogeneous Poisson point process (PPP) $\boldsymbol{\Psi} \subset \mathbb{R}^2$ with intensity $\lambda$. The IoT devices are spatially distributed according to an independent homogeneous PPP  $\boldsymbol{\Phi}\subset \mathbb{R}^2$ with intensity $\mu$. Closest BS association is assumed, where each device is served via its geographically nearest BS. A pictorial illustration of a network realization with the closest BS association is shown in Fig. \ref{Network}.  Large-scale distance-dependent signal attenuation is captured by the unbounded path-loss propagation model $r^{-\eta}$, where $r$ is the propagation distance and $\eta>2$ is the path-loss exponent.\footnote{The utilized unbounded path loss model is verified in \cite{5226961} for the aggregate network power received at a typical point for path loss exponents less than or equal to 4.} Moreover, signals are impaired by Rayleigh small-scale fading with  a  unit  mean  exponentially  distributed channel power gains,  denoted  as  ($h$). All channel gains are assumed to be independent and identically distributed irrespective of the device location and also independently from the time index. During uplink transmission, each device employs full path-loss inversion power control with target power level $\rho$ \cite{sesia2009lte}. For a device located $r_\circ$ meters away from its serving (i.e., geographically closest) BS, the transmission power of this device is given by $P_T=\rho r_\circ^{\eta}$. Consequently, the average signal power received at its serving {\rm BS} is equal to $\rho$. The target power level $\rho$ is a design parameter that is conveyed to the IoT devices via downlink signaling.

\subsection{MAC Layer and Temporal Parameters}

From the temporal perspective, the network operates according to a synchronized discrete-time system with a time slot duration of $T_{\text{\rm s}}$. Each device may generate only one data packet in its buffer at each time slot. The packets generation is modeled by a homogeneous geometric inter-arrival process with parameter $a \in [0,1]$ (packet/slot). Devices with non-empty buffers may use a time slot for an uplink transmission attempt of a single data packet. Hence, at each time slot, only one packet may arrive and/or depart the data buffer of each device in the network. Each device has a buffer that can store a maximum of $M$ data packets. Each packet is stored in the buffer until successful transmission. Each device always attempts to send the packet at the head of its data buffer, and hence, it follows a First Come First Served (FCFS) discipline. { When the device has a full buffer, newly generated packets are dropped, i.e., packet loss occurs.}

\subsection{GF-UL Transmission}

{In the proposed GF-UL scheme, we assume an opportunistic channel-aware transmission scheme. That is, only the devices that have a channel gain greater than a threshold $\tau$ can transmit \cite{5226963, 6807718}. For the exponential channel gain $\mathtt{h}$, the probability ($\Omega$) of having a channel gain greater than $\tau$ is obtained through the Complementary Cumulative Distribution Function (CCDF) of $\mathtt{h}$ as { $\Omega = 1- {F}_{\mathtt{h}}(\tau)=\exp(-\mathtt{\tau})$}. It is worth noting that the transmission threshold is a design parameter that should balance a trade-off between delay, interference, and efficient utilization of the harvested energy. From the device side, a high value of $\tau$ bias the devices to remain in the energy harvesting phase even when they have packets, which due to the high required channel gain. Hence, the devices that access the channel have two advantages: i) they are guaranteed to have good fading (i.e., greater than $\tau$) towards their serving BSs and ii) to see less number of interfering devices because of the high value of $\tau$ required for channel access, which will lead to less mutual network-wide interference. However, on the negative side, a high value of $\tau$ may also lead to unnecessary transmission deferrals that increase packet delay.}

 { By virtue of the GF-UL, {\em Tx-eligible} devices (hereafter Tx-eligible is used to denote the devices  with channel gains greater than $\tau$) directly transmit to their serving BSs without a  scheduling grant. The available bandwidth is divided into $n_{\text{\rm c}}$ orthogonal frequency resources that are universally reused over all BSs. To mitigate mutual interference, each of the Tx-eligible devices that has non-empty buffer randomly and uniformly selects one of the frequency resources for its transmission attempt. An SINR capture model is utilized, in which the received SINR has to be greater than a predefined threshold ($\theta$) for correct reception. Because of the uncoordinated nature of the GF-UL, there could be several conflicting devices simultaneously transmitting over the same resource to the same BS (i.e., intra-cell interference). Following the RF capture model~\cite{329345, Cardieri}, out of all conflicting devices transmissions, the BS can only decode the dominating one (i.e., the packet of the device with the highest received signal power) if and only if its SINR is larger than the threshold ($\theta$)}. Upon transmission success, an acknowledgment is sent to the intended device over an error-free feedback channel instantaneously.

\subsection{Battery \& Energy Harvesting Model}

\begin{figure}[t!]
	\begin{center}
		\includegraphics[width=3 in ]{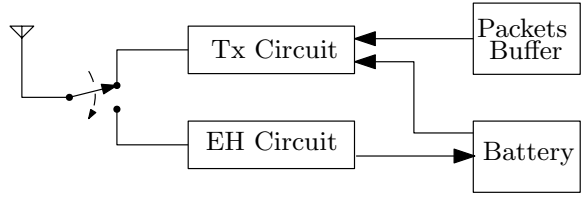}
	\end{center}
	\caption{$\;\;$ A device model that is equipped with a battery, a packets buffer, a Transmission (Tx) circuit, Energy Harvesting {(EH)} circuit, and a single antenna.}\label{battery}
\end{figure}

Each device is equipped with a rechargeable battery of capacity $B\!=\! {P}_{\text{max}} T_s$ {Watt-s} and relies on harvesting RF energy from the aggregate downlink power to recharge the battery. Hence, each device has an RF-to-DC converter with efficiency $\zeta \leq 1$. As shown in Fig.~\ref{battery}, it is assumed that the IoT devices have a single antenna, and hence, the devices cannot harvest energy and transmit at the same time. Consequently, the devices follow the ``\emph{harvest-then-transmit}'' strategy. As such, the devices keep harvesting and storing energy at their batteries until they fulfill the required energy to perform full path-loss inversion power control. {Hereafter, devices with sufficient stored energy to invert their path-loss are denoted as {\em $P_T$-capable} devices. According to the employed opportunistic GF-UL transmissions with energy harvesting, only Tx-eligible (i.e., with channel gains greater that $\tau$) and $P_T$-capable (i.e., with sufficient stored energy) devices can transmit. }  
	
{The energy harvesting and data transmissions processes are performed as follows. Devices with empty buffer operate in the harvesting mode until their battery is full. Devices with non-empty buffers switch to the transmission mode if and only if they are Tx-eligible and $P_T$-capable. Otherwise, if either of the conditions (i.e., for being Tx-eligible or $P_T$-capable) is not satisfied, then the devices operate in the harvesting mode even if they have non-empty buffers. The devices that require a transmit power greater than ${P}_{\text{max}}$ to invert their path-loss experience an outage due to the insufficient energy.}

\subsection{Methodology of Analysis}\label{bubsec:Meth_ana}

{Due to the path-loss inversion power control, the received powers at all serving BSs are independent of the devices' locations. In this case, the spatially averaged SINR performance of the typical device is a good representation for all devices in the network~\cite{Meta_Elsawy, Meta_Haenggi_Control}. On the other hand, the energy harvesting and depletion rates are location-dependent. Due to the employed downlink energy harvesting and path-loss inversion uplink power control, devices that are closer to their serving BS harvest (consume) energy at a higher (lower) rate. This results from the fixed transmission powers of the BSs and the dominant contribution of the downlink power of the serving BS to the harvested energy. Furthermore, devices that are closer to their serving BS have a lower energy depletion rate due to the smaller path-loss that needs to be inverted via the power control in each transmission attempt. To tackle the location-dependent energy harvesting and depletion rates, we divide the devices into $n\in \{1,2, \cdots, N\}$ equiprobable location-dependent energy-harvesting (EH)-classes based on distances to their serving BSs.}

\begin{figure*}[t!]
	\centering
	\begin{subfigure}[t]{0.99\textwidth}
		\centerline{\begin{tikzpicture}[fill=blue,ultra thick, scale = .55, transform shape,font=\small]

 \definecolor{red}{rgb}{1,0.27,0}
\definecolor{mygreen}{rgb}{0.20, 0.8, 0.2}
\definecolor{myblue}{rgb}{0, 0, 0.8}
\definecolor{myorangeange}{RGB}{255, 178, 102}
\definecolor{mymagenta}{rgb}{0.78, 0.08, 0.52}
\definecolor{mycyan}{rgb}{0, 0.74, 1}
\definecolor{depc}{rgb}{0.12, 0.3, 0.17}
\definecolor{upc}{rgb}{0.8, 0.0, 0.0}
\definecolor{downc}{rgb}{0.12, 0.3, 0.17}
\definecolor{samec}{rgb}{0.0, 0.75, 1.0}

\tikzset{
    /pgf/arrow keys/second color/.store in=\mycolor,
    /pgf/arrow keys/second color=green
}

\tikzset{->-/.style={decoration={   markings,  mark=at position .5 with {\arrow[thick]{stealth}}},postaction={decorate}}}

\node (Id) at (0,0) [circle, minimum width=1cm, font=\normalsize,align=center] { {}};

\node (L1) at ($(Id)+(-8cm,0cm)$) [circle, draw, line width=0.75pt, rounded corners=1ex, fill=mycyan!60!white, minimum width=1.8cm, font=\normalsize,align=center] { $0$};
\node (L2) at ($(L1)+(3.5cm,0cm)$) [circle, draw, line width=0.75pt, rounded corners=1ex, fill=mycyan!60!white, minimum width=1.8cm, font=\normalsize,align=center] { $1$};
\node (L3) at ($(L2)+(+3.5cm,0cm)$) [circle, draw, line width=0.75pt, rounded corners=1ex, fill=mycyan!60!white, minimum width=1.8cm, font=\normalsize,align=center] { $2$};
\node (L4) at ($(L3)+(+3.5cm,0cm)$) [circle, line width=0.75pt, rounded corners=1ex, minimum width=1.8cm, font=\Huge,align=center] {\dots};
\node (L5) at ($(L4)+(+3.5cm,0cm)$) [circle, draw, line width=0.75pt, rounded corners=1ex, fill=mycyan!70!white, minimum width=1.8cm, font=\normalsize,align=center] { $M-1$};
\node (L6) at ($(L5)+(+3.5cm,0cm)$) [circle, draw, line width=0.75pt, rounded corners=1ex, fill=mycyan!70!white, minimum width=1.8cm, font=\normalsize,align=center] { $M$};

\draw [->-, myblue,  line width=0.6pt, bend left=50]  (L1)  to node [above, rotate=0, font=\small , color=black,pos=0.5,xshift=0cm,yshift=0cm, rotate=0] {$a \mathbf{H}^{[n]}$}  (L2);
\draw [->-, myblue,  line width=0.6pt, bend left=50]  (L2)  to node [above, rotate=0, font=\small , color=black,pos=0.5,xshift=0cm,yshift=0cm, rotate=0] {$a (\Omega \mathbf{F}^{[n]} + \bar{\Omega} \mathbf{H}^{[n]}) $}  (L3);
\draw [->-, myblue,  line width=0.6pt, bend left=50]  (L3)  to node [above, rotate=0, font=\small , color=black,pos=0.5,xshift=0cm,yshift=0cm, rotate=0] {$a (\Omega \mathbf{F}^{[n]} + \bar{\Omega} \mathbf{H}^{[n]})  $}  (L4);
\draw [->-, myblue,  line width=0.6pt, bend left=50]  (L4)  to node [above, rotate=0, font=\small , color=black,pos=0.5,xshift=0cm,yshift=0cm, rotate=0] {$a (\Omega \mathbf{F}^{[n]} + \bar{\Omega} \mathbf{H}^{[n]})  $}  (L5);
\draw [->-, myblue,  line width=0.6pt, bend left=50]  (L5)  to node [above, rotate=0, font=\small , color=black,pos=0.5,xshift=0cm,yshift=0cm, rotate=0] {$a (\Omega \mathbf{F}^{[n]} + \bar{\Omega} \mathbf{H}^{[n]})  $}  (L6);

\draw [->-, mygreen,  line width=0.6pt, bend left=50]  (L6)  to node [below, rotate=0, font=\small , color=black,pos=0.5,xshift=0cm,yshift=0cm, rotate=0] {$\bar{a}{\Omega}\mathbf{S}^{[n]} $}  (L5);
\draw [->-, mygreen,  line width=0.6pt, bend left=50]  (L5)  to node [below, rotate=0, font=\small , color=black,pos=0.5,xshift=0cm,yshift=0cm, rotate=0] {$\bar{a}{\Omega}\mathbf{S}^{[n]}$}  (L4);
\draw [->-, mygreen,  line width=0.6pt, bend left=50]  (L4)  to node [below, rotate=0, font=\small , color=black,pos=0.5,xshift=0cm,yshift=0cm, rotate=0] {$\bar{a}{\Omega}\mathbf{S}^{[n]}$}  (L3);
\draw [->-, mygreen,  line width=0.6pt, bend left=50]  (L3)  to node [below, rotate=0, font=\small , color=black,pos=0.5,xshift=0cm,yshift=0cm, rotate=0] {$\bar{a}{\Omega}\mathbf{S}^{[n]}$}  (L2);
\draw [->-, mygreen,  line width=0.6pt, bend left=50]  (L2)  to node [below, rotate=0, font=\small , color=black,pos=0.5,xshift=0cm,yshift=0cm, rotate=0] {$\bar{a}{\Omega}\mathbf{S}^{[n]}$}  (L1);

\draw[->-,dashed, mymagenta,  line width=0.65pt,] (L1) to [pos=0.5, minimum size=0.5mm, out=250,in=300,loop,looseness=6]  (L1);
\node (l1) at ($(L1)-(0cm,2.5cm)$) [font=\small,align=center, rotate=0] {$\bar{a} \mathbf{H}^{[n]}$};

\draw[->-,dashed, mymagenta,  line width=0.65pt,] (L2) to [pos=0.5, minimum size=0.5mm, out=250,in=300,loop,looseness=6]  (L2);
\node (l2) at ($(L2)-(0cm,2.8cm)$) [font=\small,align=center, rotate=0] {$\Omega  ( \bar{a} \mathbf{F}^{[n]} + {a} \mathbf{S}^{[n]})$ \\ $+\bar{\Omega} \bar{a}  \mathbf{H}^{[n]} $};

\draw[->-,dashed, mymagenta,  line width=0.65pt,] (L3) to [pos=0.1, minimum size=0.5mm, out=250,in=300,loop,looseness=6]  (L3);
\node (l3) at ($(L3)-(0cm,2.8cm)$) [font=\small,align=center, rotate=0] {$\Omega  ( \bar{a} \mathbf{F}^{[n]} + {a} \mathbf{S}^{[n]})$ \\ $+\bar{\Omega} \bar{a}  \mathbf{H}^{[n]} $};

\draw[->-,dashed, mymagenta,  line width=0.65pt,] (L5) to [pos=0.5, minimum size=0.5mm, out=250,in=300,loop,looseness=6]  (L5);
\node (l5) at ($(L5)-(0cm,2.8cm)$) [font=\small,align=center, rotate=0] {$\Omega  ( \bar{a} \mathbf{F}^{[n]} + {a} \mathbf{S})$ \\ $+\bar{\Omega} \bar{a}  \mathbf{H}^{[n]} $};

\draw[->-,dashed, mymagenta,  line width=0.65pt,] (L6) to [pos=0.5, minimum size=0.5mm, out=250,in=300,loop,looseness=6]  (L6);
\node (l6) at ($(L6)-(0cm,2.8cm)$) [font=\small,align=center, rotate=0] {$\Omega  ( \mathbf{1} - {a} \mathbf{S}^{[n]})$ \\ $+\bar{\Omega} \bar{a}  \mathbf{H}^{[n]}  $};

\end{tikzpicture}}
		\caption{ Data buffer states}
		\label{fig:MC_buf}
	\end{subfigure}
~
	\begin{subfigure}[t]{0.99\textwidth}
		\centerline{\begin{tikzpicture}[fill=blue,ultra thick, scale = .55, transform shape,font=\small]

 \definecolor{red}{rgb}{1,0.27,0}
\definecolor{mygreen}{rgb}{0.20, 0.8, 0.2}
\definecolor{myblue}{rgb}{0, 0, 0.8}
\definecolor{myorangeange}{RGB}{255, 178, 102}
\definecolor{mymagenta}{rgb}{0.78, 0.08, 0.52}
\definecolor{mycyan}{rgb}{0, 0.74, 1}
\definecolor{depc}{rgb}{0.12, 0.3, 0.17}
\definecolor{upc}{rgb}{0.8, 0.0, 0.0}
\definecolor{downc}{rgb}{0.12, 0.3, 0.17}
\definecolor{samec}{rgb}{0.0, 0.75, 1.0}

\tikzset{
    /pgf/arrow keys/second color/.store in=\mycolor,
    /pgf/arrow keys/second color=green
}

\tikzset{->-/.style={decoration={   markings,  mark=at position .5 with {\arrow[thick]{stealth}}},postaction={decorate}}}

\node (Id) at (0,0) [circle, minimum width=1cm, font=\normalsize,align=center] { {}};




\node (b2) at (0.5,4) [preaction={fill=mycyan!60!white,fill opacity=0.6},rounded corners=1ex, minimum width=22cm, minimum height = 3cm, font=\small\bfseries,align=left] {};


\node (C1) at ($(Id)+(-3.5 cm,0cm)$) [circle, draw, line width=0.5pt, rounded corners=1ex, fill=mygreen!60!white, minimum width=0.8cm, font=\normalsize,align=left] {};
\node (C11) at ($(C1)+(2.5cm,0cm)$) [font=\normalsize,align=left] {EH states  with \\  overall probability $\bar{\delta}$};

\node (C12) at ($(C1)+(5.5 cm,0cm)$) [circle, draw, line width=0.5pt, rounded corners=1ex, fill=red!70!white, minimum width=0.8cm, font=\normalsize,align=left] {};
\node (C122) at ($(C12)+(2.7cm,-0.05cm)$) [font=\normalsize,align=left] {$P_T$-capable states with \\  overall probability $\delta$};

\node (l2) at (-10,4) [font=\Large ,align=center, rotate=90] {$m>0 $};





\node (L1) at ($(Id)+(-8cm,4cm)$) [circle, draw, line width=0.75pt, rounded corners=1ex, fill=mygreen!60!white, minimum width=1.8cm, font=\normalsize,align=center] { $0$};
\node (L2) at ($(L1)+(3.5cm,0cm)$) [circle, draw, line width=0.75pt, rounded corners=1ex, fill=mygreen!60!white, minimum width=1.8cm, font=\normalsize,align=center] { $1$};
\node (L3) at ($(L2)+(+3.5cm,0cm)$) [circle, draw, line width=0.75pt, rounded corners=1ex, fill=mygreen!60!white, minimum width=1.8cm, font=\normalsize,align=center] { $2$};
\node (L4) at ($(L3)+(+3.5cm,0cm)$) [circle, draw, line width=0.75pt, rounded corners=1ex, fill=red!70!white, minimum width=1.8cm, font=\normalsize,align=center] { $3$};
\node (L5) at ($(L4)+(+3.5cm,0cm)$) [circle, draw, line width=0.75pt, rounded corners=1ex, fill=red!70!white, minimum width=1.8cm, font=\normalsize,align=center] { $4$};
\node (L6) at ($(L5)+(+3.5cm,0cm)$) [circle, draw, line width=0.75pt, rounded corners=1ex, fill=red!70!white, minimum width=1.8cm, font=\normalsize,align=center] { $5$};

\draw [->-, dashed, myblue,  line width=0.6pt, bend left=20]  (L1)  to node [above, rotate=0, font=\small , color=black,pos=0.5,xshift=0cm,yshift=0cm, rotate=0] {$p_1$}  (L2);
\draw [->-, dashed, myblue,  line width=0.6pt, bend left=20]  (L2)  to node [above, rotate=0, font=\small , color=black,pos=0.5,xshift=0cm,yshift=0cm, rotate=0] {$p_1$}  (L3);
\draw [->-, dashed, myblue,  line width=0.6pt, bend left=20]  (L3)  to node [above, rotate=0, font=\small , color=black,pos=0.5,xshift=0cm,yshift=0cm, rotate=0] {$p_1$}  (L4);
\draw [->-, dashed, myblue,  line width=0.6pt, bend left=20]  (L4)  to node [above, rotate=0, font=\small , color=black,pos=0.5,xshift=0cm,yshift=0cm, rotate=0] {$\bar{\Omega} p_1$}  (L5);
\draw [->-, dashed, myblue,  line width=0.6pt, bend left=20]  (L5)  to node [above, rotate=0, font=\small , color=black,pos=0.5,xshift=0cm,yshift=0cm, rotate=0] {$\bar{\Omega} \bar{p}_0$}  (L6);

\draw [->-, dashed, myblue,  line width=0.6pt, bend left=50]  (L1)  to node [below, rotate=0, font=\small , color=black,pos=0.5,xshift=0cm,yshift=0cm, rotate=0] {$p_2$}  (L3);
\draw [->-, dashed, myblue,  line width=0.6pt, bend left=50]  (L2)  to node [below, rotate=0, font=\small , color=black,pos=0.5,xshift=0cm,yshift=0cm, rotate=0] {$p_2$}  (L4);

\draw [->-, dashed, myblue,  line width=0.6pt, bend left=55]  (L3)  to node [below, rotate=0, font=\small , color=black,pos=0.5,xshift=0cm,yshift=0cm, rotate=0] {$p_2$}  (L5);
\draw [->-, dashed, myblue,  line width=0.6pt, bend left=55]  (L4)  to node [below, rotate=0, font=\small , color=black,pos=0.5,xshift=0.3cm,yshift=-0.1cm, rotate=0] {$\bar{\Omega} \overline{(p_0+p_1)}$}  (L6);

\draw [->-, dashed, myblue,  line width=0.6pt, bend left=60]  (L1)  to node [above, rotate=0, font=\small , color=black,pos=0.5,xshift=0cm,yshift=0cm, rotate=0] {$p_3$}  (L4);
\draw [->-, dashed, myblue,  line width=0.6pt, bend left=60]  (L2)  to node [above, rotate=0, font=\small , color=black,pos=0.5,xshift=0cm,yshift=0cm, rotate=0] {$p_3$}  (L5);
\draw [->-, dashed, myblue,  line width=0.6pt, bend left=60]  (L3)  to node [above, rotate=0, font=\small , color=black,pos=0.5,xshift=0.5cm,yshift=0cm, rotate=0] {$\overline{(p_0+p_1+p_2)}$}  (L6);

\draw [->-, dashed, myblue,  line width=0.6pt, bend left=70]  (L1)  to node [above, rotate=0, font=\small , color=black,pos=0.5,xshift=0cm,yshift=0cm, rotate=0] {$p_4$}  (L5);
\draw [->-, dashed, myblue,  line width=0.6pt, bend left=70]  (L2)  to node [above, rotate=0, font=\small , color=black,pos=0.5,xshift=0.5cm,yshift=0cm, rotate=0] {$\overline{(p_0+p_1+p_2+p_3)}$}  (L6);

\draw [->-, dashed, myblue,  line width=0.6pt, bend left=80]  (L1)  to node [above, rotate=0, font=\small , color=black,pos=0.5,xshift=0cm,yshift=0cm, rotate=0] {$\overline{(p_0+p_1+p_2+p_3+p_4)}$}  (L6);

\draw [->-, dashed, red,  line width=0.6pt, bend left=40]  (L4)  to node [below, rotate=0, font=\small , color=black,pos=0.5,xshift=0cm,yshift=0cm, rotate=0] {$\Omega$}  (L1);
\draw [->-, dashed, red,  line width=0.6pt, bend left=40]  (L5)  to node [below, rotate=0, font=\small , color=black,pos=0.5,xshift=0cm,yshift=0cm, rotate=0] {$\Omega$}  (L2);
\draw [->-, dashed, red,  line width=0.6pt, bend left=40]  (L6)  to node [below, rotate=0, font=\small , color=black,pos=0.5,xshift=0cm,yshift=0cm, rotate=0] {$\Omega$}  (L3);

\draw[->-, mymagenta,  line width=0.65pt,] (L1) to [pos=0.5, minimum size=0.5mm, out=200,in=270,loop,looseness=2]  (L1);
\node (l1) at ($(L1)-(1cm,1.3cm)$) [font=\small,align=center, rotate=0] {$p_0$};
\draw[->-, mymagenta,  line width=0.65pt,] (L2) to [pos=0.5, minimum size=0.5mm, out=200,in=270,loop,looseness=2]  (L2);
\node (l2) at ($(L2)-(1cm,1.3cm)$) [font=\small,align=center, rotate=0] {$p_0$};
\draw[->-, mymagenta,  line width=0.65pt,] (L3) to [pos=0.5, minimum size=0.5mm, out=200,in=270,loop,looseness=2]  (L3);
\node (l3) at ($(L3)-(1cm,1.3cm)$) [font=\small,align=center, rotate=0] {$p_0$};
\draw[->-, mymagenta,  line width=0.65pt,] (L4) to [pos=0.5, minimum size=0.5mm, out=340,in=270,loop,looseness=2]  (L4);
\node (l3) at ($(L4)-(-0cm,1.4cm)$) [font=\small,align=center, rotate=0] {$\bar{\Omega} p_0$};
\draw[->-, mymagenta,  line width=0.65pt,] (L5) to [pos=0.5, minimum size=0.5mm, out=340,in=270,loop,looseness=2]  (L5);
\node (l3) at ($(L5)-(-0cm,1.4cm)$) [font=\small,align=center, rotate=0] {$\bar{\Omega} p_0$};

\draw[->-, mymagenta,  line width=0.65pt,] (L6) to [pos=0.5, minimum size=0.5mm, out=340,in=270,loop,looseness=2]  (L6);
\node (l3) at ($(L6)-(-1cm,1.4cm)$) [font=\small,align=center, rotate=0] {$\bar{\Omega}$};

\end{tikzpicture}}
		\caption{ The underlying battery states for a device with non-empty buffer for $L=5$ and $d^{[n]}=3$. }
		\label{fig:MC_bat}
	\end{subfigure}
	\caption{$\;\;$ The Markov models for the data buffers and battery  where $\delta$ is the probability to be $P_T$-capable, $\Omega$ is the probability to be Tx-eligible, $a$ is the probability of packet generation, $p_c$ is the probability of successful transmission, and $p_l$ is the probability of harvesting $l$ energy units in one time slot.}
	\label{fig:MC_sep}
\end{figure*}
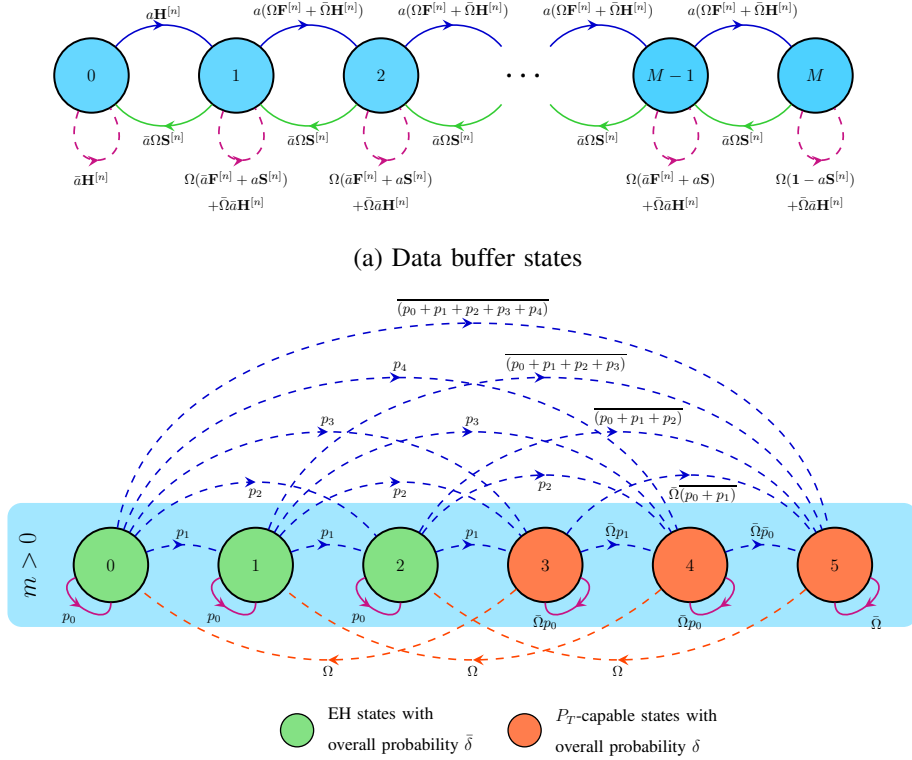

\begin{figure}[t!]
	\centering
	\begin{tikzpicture}[fill=blue,ultra thick, scale = .55, transform shape,font=\small]

 \definecolor{red}{rgb}{1,0.27,0}
\definecolor{mygreen}{rgb}{0.20, 0.8, 0.2}
\definecolor{myblue}{rgb}{0, 0, 0.8}
\definecolor{myorangeange}{RGB}{255, 178, 102}
\definecolor{mymagenta}{rgb}{0.78, 0.08, 0.52}
\definecolor{mycyan}{rgb}{0, 0.74, 1}
\definecolor{depc}{rgb}{0.12, 0.3, 0.17}
\definecolor{upc}{rgb}{0.8, 0.0, 0.0}
\definecolor{downc}{rgb}{0.12, 0.3, 0.17}
\definecolor{samec}{rgb}{0.0, 0.75, 1.0}

\tikzset{
    /pgf/arrow keys/second color/.store in=\mycolor,
    /pgf/arrow keys/second color=green
}

\tikzset{->-/.style={decoration={   markings,  mark=at position .5 with {\arrow[thick]{stealth}}},postaction={decorate}}}

\node (Id) at (0,0) [circle, minimum width=1cm, font=\normalsize,align=center] { {}};



\node (b1) at (-0.5,4) [preaction={fill=mycyan!60!white,fill opacity=0.6},rounded corners=1ex, minimum width=15cm, minimum height = 2.2cm, font=\Large\bfseries,align=left] {};
\node (b2) at (-0.5,8) [preaction={fill=mycyan!60!white,fill opacity=0.6},rounded corners=1ex, minimum width=15cm, minimum height = 2.2cm, font=\Large\bfseries,align=left] {};
\node (b3) at (-0.5,12)[preaction={fill=mycyan!60!white,fill opacity=0.6},rounded corners=1ex, minimum width=15cm, minimum height = 2.2cm, font=\Large\bfseries,align=left] {};
\node (b4) at (-0.5,16)[preaction={fill=mycyan!60!white,fill opacity=0.6},rounded corners=1ex, minimum width=15cm, minimum height = 2.2cm, font=\Large\bfseries,align=left] {};


\node (C1) at ($(Id)+(-2cm,2cm)$) [circle, draw, line width=0.5pt, rounded corners=1ex, fill=mygreen!60!white, minimum width=0.1cm, font=\small,align=center] {};
\node (C11) at ($(C1)+(1.1cm,0cm)$) [font=\small,align=center] {EV states};

\node (C12) at ($(C1)+(3 cm,0cm)$) [circle, draw, line width=0.5pt, rounded corners=1ex, fill=red!70!white, minimum width=0.1cm, font=\small,align=center] {};
\node (C122) at ($(C12)+(1.7cm,0cm)$) [font=\small,align=center] {$P_T$=capable states};

\node (l1) at (-7.5,4) [font=\Large,align=center, rotate=90] {$m =0$};
\node (l2) at (-7.5,8) [font=\Large,align=center, rotate=90] {$m = 1 $};
\node (l3) at (-7.5,12) [font=\Large,align=center, rotate=90] {$ m=2$};
\node (l4) at (-7.5,16) [font=\Large,align=center, rotate=90] {$ m= M$};





\node (L11) at ($(Id)+(-6cm,4cm)$) [circle, draw, line width=0.75pt, rounded corners=1ex, fill=mygreen!60!white, minimum width=1.4cm, font=\normalsize,align=center] { $0,0$};
\node (L12) at ($(L11)+(3.2cm,0cm)$) [circle, draw, line width=0.75pt, rounded corners=1ex, fill=mygreen!60!white, minimum width=1.4cm, font=\normalsize,align=center] { $0,1$};
\node (L13) at ($(L12)+(2.4cm,0cm)$) [ font=\huge   ,align=center]{\textcolor{black}{$\dots$ }};

\node (L21) at ($(Id)+(+6cm,4cm)$) [circle, draw, line width=0.75pt, rounded corners=1ex, fill=mygreen!60!white, minimum width=1.4cm, font=\normalsize,align=center] { $0,L$};
\node (L22) at ($(L21)+(-2.1cm,0cm)$) [ font=\huge   ,align=center]{\textcolor{black}{$\dots$ }};
\node (L23) at ($(L22)+(-2.3cm,0cm)$) [circle, draw, line width=0.75pt, rounded corners=1ex, fill=mygreen!60!white, minimum width=1.4cm, font=\normalsize,align=center] { $0,d^{[n]}$};

\node (LL11) at ($(Id)+(-6cm,8cm)$) [circle, draw, line width=0.75pt, rounded corners=1ex, fill=mygreen!60!white, minimum width=1.4cm, font=\normalsize,align=center] { $1,0$};
\node (LL12) at ($(LL11)+(3.2cm,0cm)$) [circle, draw, line width=0.75pt, rounded corners=1ex, fill=mygreen!60!white, minimum width=1.4cm, font=\normalsize,align=center] { $1,1$};
\node (LL13) at ($(LL12)+(2.4cm,0cm)$) [ font=\huge   ,align=center]{\textcolor{black}{$\dots$ }};

\node (LL21) at ($(Id)+(+6cm,8cm)$) [circle, draw, line width=0.75pt, rounded corners=1ex, fill=red!70!white, minimum width=1.4cm, font=\normalsize,align=center] { $1,L$};
\node (LL22) at ($(LL21)+(-2.1cm,0cm)$) [ font=\huge   ,align=center]{\textcolor{black}{$\dots$ }};
\node (LL23) at ($(LL22)+(-2.3cm,0cm)$) [circle, draw, line width=0.75pt, rounded corners=1ex, fill=red!70!white, minimum width=1.4cm, font=\normalsize,align=center] { $1,d^{[n]}$};

\node (Lm11) at ($(Id)+(-6cm,12cm)$) [circle, draw, line width=0.75pt, rounded corners=1ex, fill=mygreen!60!white, minimum width=1.4cm, font=\normalsize,align=center] { $2,0$};
\node (Lm12) at ($(Lm11)+(3.2cm,0cm)$) [circle, draw, line width=0.75pt, rounded corners=1ex, fill=mygreen!60!white, minimum width=1.4cm, font=\normalsize,align=center] { $2,1$};
\node (Lm13) at ($(Lm12)+(2.4cm,0cm)$) [ font=\huge   ,align=center]{\textcolor{black}{$\dots$ }};

\node (Lm21) at ($(Id)+(+6cm,12cm)$) [circle, draw, line width=0.75pt, rounded corners=1ex, fill=red!70!white, minimum width=1.4cm, font=\normalsize,align=center] { $2,L$};
\node (Lm22) at ($(Lm21)+(-2.1cm,0cm)$) [ font=\huge   ,align=center]{\textcolor{black}{$\dots$ }};
\node (Lm23) at ($(Lm22)+(-2.3cm,0cm)$) [circle, draw, line width=0.75pt, rounded corners=1ex, fill=red!70!white, minimum width=1.4cm, font=\normalsize,align=center] { $2,d^{[n]}$};

\node (LM11) at ($(Id)+(-6cm,16cm)$) [circle, draw, line width=0.75pt, rounded corners=1ex, fill=mygreen!60!white, minimum width=1.4cm, font=\normalsize,align=center] { $M,0$};
\node (LM12) at ($(Lm11)+(3.2cm,4cm)$) [circle, draw, line width=0.75pt, rounded corners=1ex, fill=mygreen!60!white, minimum width=1.4cm, font=\normalsize,align=center] { $M,1$};
\node (LM13) at ($(Lm12)+(2.4cm,4cm)$)[ font=\huge   ,align=center]{\textcolor{black}{$\dots$ }};

\node (LM21) at ($(Id)+(+6cm,16cm)$) [circle, draw, line width=0.75pt, rounded corners=1ex, fill=red!70!white, minimum width=1.4cm, font=\normalsize,align=center] { $M,L$};
\node (LM22) at ($(Lm21)+(-2.1cm,4cm)$)[ font=\huge   ,align=center]{\textcolor{black}{$\dots$ }};
\node (LM23) at ($(Lm22)+(-2.3cm,4cm)$) [circle, draw, line width=0.75pt, rounded corners=1ex, fill=red!70!white, minimum width=1.4cm, font=\normalsize,align=center] { $M,d^{[n]}$};

\node (d11) at  ($(Lm11)+(0cm,2cm)$)[ font=\huge  ,align=center]{\textcolor{black}{$\vdots$ }};
\node (d12) at ($(Lm12)+(0cm,2cm)$)[ font=\huge  ,align=center]{\textcolor{black}{$\vdots$ }};
\node (d13) at ($(Lm13)+(0cm,2cm)$)[ font=\huge   ,align=center]{\textcolor{black}{$\vdots$ }};
\node (d21) at ($(Lm21)+(0cm,2cm)$)[ font=\huge   ,align=center]{\textcolor{black}{$\vdots$ }};
\node (d22) at ($(Lm22)+(0cm,2cm)$)[ font=\huge   ,align=center]{\textcolor{black}{$\vdots$ }};
\node (d23) at  ($(Lm23)+(0cm,2cm)$)[ font=\huge   ,align=center]{\textcolor{black}{$\vdots$ }};

\end{tikzpicture}
	\vspace{-5mm}
	\caption{  $\;\;$Two-dimensional DTMC for a test device in the $n^{th}$ class. The green color states indicate the energy harvesting states and the red color states indicate  the the transmitting states. }\label{fig_2D_DTMC}
\end{figure}
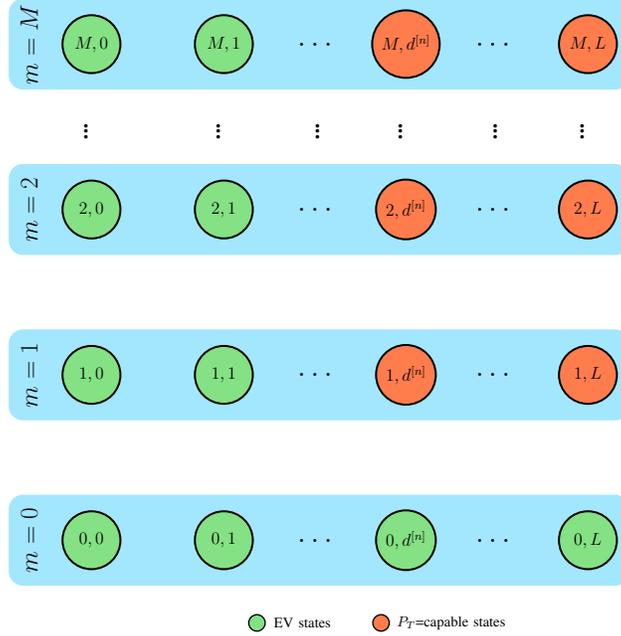
{To account for the joint states of the data buffer and batteries of each device, we discretize the battery into $(L+1)$ energy levels of equal amount $w$. That is, the total battery capacity is given by $B=Lw$. The discretization of the battery levels enables a simple DTMC representation for the temporal evolution battery states. An illustration for the individual DTMC for the data buffer and the underlying DTMC of the battery states are shown in  Fig.~\ref{fig:MC_sep}. For the sake of analysis, the data buffer and battery state should be jointly considered via the two-dimensional (2D) DTMC shown in Fig.~\ref{fig_2D_DTMC}, where the state $(m,l)$ indicates that the device has $m$ packets in its buffer and $l$ energy units in its battery. Note that the transition probabilities among the states are not shown in the Fig.~\ref{fig_2D_DTMC} to avoid overcrowded exposition. However, it is worth noting the following
\begin{itemize}
	\item The DTMC shown in Fig.~\ref{fig_2D_DTMC} is a representation of the joint states of the data buffer and battery of each device in the network.
	\item The state transition probabilities 2D DTMC can be deduced from the two DTMCs of Fig.~\ref{fig:MC_sep}.  
	\item The levels $m\in \{0,1,2,\cdots,M\}$ track the number of packets in data buffer and the phases  $l\in \{0,1,2,\cdots,L\}$ track the number of energy units in battery. 
		\item The green states denote the energy harvesting mode.
		\item The orange states denote sufficient stored energy for transmissions. A device in any of the orange states switches to the transmission mode if Tx-eligible (i.e., with probability $\Omega$) and stays in the energy harvesting mode otherwise. 
		\item The green and orange states are discriminated according to the number of energy units required for one transmission attempt, which is given by $d^{[n]}= \ceil{\frac{\mathcal{P}_T^{[n]}}{w}}$, where  $\mathcal{P}_T^{[n]}$ is the energy required for one uplink transmission attempt. 
		\item A transmitting device with battery state $l$ goes to state $ l'= \floor{l- \frac{\mathcal{P}_T^{[n]}}{w}}=l-d^{[n]}$. 
		\item A harvesting device at state $l$ goes to state $l'= \floor{l + \frac{\mathcal{P}^{[n]}_H}{w}}$, where $\mathcal{P}_H^{[n]} $ is the aggregate downlink harvested energy after the RF-to-DC converter within one time slot. 
		\item A transmission attempt is successful with probability $p_c$, which is the probability that the transmission dominates all other inter-cell transmissions and have an SINR greater than $\theta$.
		\item Within one time slot, a device at state $m$ may either stay within the same state or have one step transition to state $m+1$ or state $m-1$ depending on the packet arrival and departure. 
		\item Note that ${d}^{[n]}$,  $\mathcal{P}_T^{[n]}$, and $\mathcal{P}^{[n]}_H$ are EH-class dependent. However, $p_c$ is independent of the device location due to the employed path-loss channel inversion power control.
\end{itemize}}

{From the aforementioned illustration, it is clear that the DTMCs of all devices are interdependent due to mutual interference. That is, the successful packet departure probability $p_c$ at one device depends on the mode of operation and transmission powers of all other devices in the network. To obtain $p_c$ for a given device, the states probabilities of the DTMC for all other devices are required. Meanwhile, $p_c$ is required to solve the DTMC and obtain the underlying state probabilities for all devices. To solve such interdependence, the following approach is applied:  (i) stochastic geometry is used to characterize the EH-classes along with the transition probabilities among all battery states for each EH-class; (ii) stochastic geometry is utilized to find $p_c$ as a function of the DTMC states probabilities $\mathbf{x}^{[n]}$ of all devices that may belong to different EH-classes; (iii) the DTMC solution for each EH-class $\mathbf{x}^{[n]}$ is obtained in terms of $p_c$; and (iv) iterate between the stochastic geometry analysis in (ii) and queueing theory analysis in (iii) until convergence, which is guaranteed by virtue of the fixed point theorem.}

\section{Performance Analysis} \label{sic:Performance_Analysis}

{An arbitrary, yet fixed, realizations for $\boldsymbol{\Psi}$ and $\boldsymbol{\Phi}$ are considered, which implies that the locations of the BSs and devices do not change over time. On the other hand, the channel gains randomly and independently change from one time slot to another. The buffer states, battery states, and devices activities randomly change over time according to the arrival/departure of packets and energy harvesting/depletion processes. Note that the fixed devices' location is a common assumption in spatio-temporal analysis due to the much smaller time scale of time slots when compared to the devices' mobility. As mentioned earlier, such fixed network topology implies location-dependent energy harvesting and depletion processes. On the other hand, the location impact on the SINR is counteracted with the path-loss inversion power control, which leads to a unified success probability $p_c$ for all devices in the network. In the sequel, we first characterize the location-dependent energy harvesting/depletion process. Then, we present the analysis for the transmission success probability. To this end, the EH-classes are defined and the DTMC analysis for each class is detailed. Finally, the iterative solution is presented.  For a quick reference, the notation used in this paper is summarized in Table~\ref{table_notation}.}

\begin{table}[t!]
	\centering
	\footnotesize
	\renewcommand{\arraystretch}{1.3}
	\begin{tabular}{||p{1.2cm}|p{10.2cm}|} 
		\hline  \textbf{Notation} & \textbf{Description} \\
		\hline $\mu$; $\lambda$ & device density; {\rm BS} density; \\
		\hline $h$; $\Omega$; $\tau$  & channel gain; transmission probability; transmission channel gain threshold   \\
		\hline $\theta$; $\rho$ & detection threshold for successful transmission;  power control parameter  \\
		\hline ${n_{c}}$ & number of orthogonal codes for uplink transmission \\	
		\hline ${p_{c}}$; $\delta$ & probability of successful transmission; probability of being in transmission states\\	
		\hline $\eta$; $\sigma^2$ & path-loss exponent; noise power  \\	
		\hline $T_s$; $\zeta$; $P$ & time slot duration; harvesting energy efficiency; BS transmitting power  \\	
		\hline $M$; $a$ & device's data buffer size; geometric arrival parameter   \\	
		\hline $B$; $L$; $w$ & device's battery size; number of energy levels; spacing between energy levels   \\		
		\hline
	\end{tabular}
	\caption{{: Summary of Notation}}\label{table_notation}
\end{table} 

\subsection{Stochastic Geometry Analysis} \label{sic:Performance_Analysis1}

Next, we characterize the harvested/depleted energy and the transmission success probability using stochastic geometry analysis as functions of the queuing parameters. 

{\subsubsection{Harvested/Depleted Energy Analysis} In this section, the energy harvesting/depletion processes are characterized in terms of the distance from the device to its serving BS, denoted as $r_\circ$. For simplicity, $r_\circ$ is discretized to $N$ equiprobable location-dependent EH-classes.\footnote{As in any quantization model for a continuous variable, increasing the number of quantization levels improves the model accuracy, with the cost of more computations. In our paper, we selected $N=50$ to provide a reasonable trade-off between accuracy and complexity.}  Given the independence between $\boldsymbol{\Psi}$ and $\boldsymbol{\Phi}$, the serving distances have a Rayleigh distribution with a CDF given by} \cite{andrews2011tractable,martin_book}
	\begin{align}
	F_{r_\circ}(r)=1-\exp\left\{- r^2  \pi \lambda\right\}, \quad 0\leq r\leq \infty. \label{r_dis}
	\end{align}
	Let $R_n$ be the minimum distance for EH-class $n$. The devices can be classified into $N$ equiprobable classes by setting $R_1=0$ and finding $R_n$ for $n>1$ as  $F_{r_\circ}(r)(R_{n+1})-F_{r_\circ}(r)(R_{n})=\frac{1}{N}$. Then, the discretized distances $r_n$ is selected as the median distance for the devices in each EH-class $n$, which is expressed as
	\begin{align}
	r_n\!=\! \sqrt{\! \frac{-\ln\left(\!1\!-\frac{F_{r_\circ}(r)(R_{n+1})+F_{r_\circ}(r)(R_n)}{2}\right)}{\pi \lambda}}, \;\; 1 \leq n \leq N.\label{mid_point}
	\end{align}
	
		\begin{figure}[t!]
		\begin{center}
			{\includegraphics[width=2.95 in ]{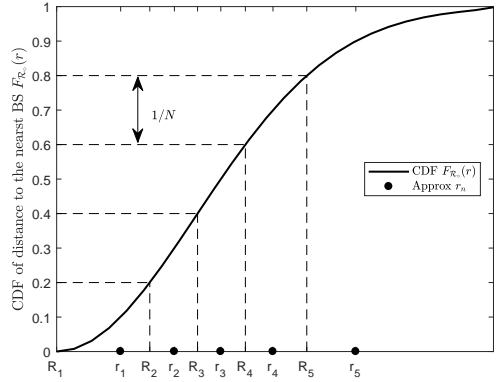}}
		\end{center}
		\caption{  $\;$ An example of distance boundaries for $N\!\!=\!5$ classes over the CDF of the distance to the nearest BS. The black dots represent the approximated distance for the devices in each class $r_n$. }\label{Distance_example}
	\end{figure}
	Equation \eqref{mid_point} follows from the inverse function of \eqref{r_dis} for the median $\frac{F_{r_\circ}(r)(R_{n+1})+F_{r_\circ}(r)(R_n)}{2}$ between the distances $R_n$ and $R_{n+1}$. Fig. \ref{Distance_example} shows an example for classifying the devices according to $N=5$ EH-classes. The  discretized distance for each EH-class $r_n$ is highlighted in Fig.\ref{Distance_example} by a black dot.	{Using the discretized distances $r_n$, the harvested/depleted energy can be characterized for each EH-class. For instance, the energy depleted for a device in EH-class $n$ in each uplink transmission attempt is given by $\mathcal{P}^{[n]}_T = T_s \rho r_n^\eta$,  where the superscript $[n]$ is used to emphasize the EH-class. In the downlink, the energy harvested per time slot for a device in EH-class $n$ can be expressed at }
\begin{align}\label{eh_1}
\mathcal{P}_H^{[n]} =& T_s \zeta P \left( {g}_\circ {r}_n^{-\eta}+ \sum\limits_{u_i \in \boldsymbol{\Psi} \setminus\{u_{\circ}\}}  {g}_i ||u_i-\ell_\circ||^{-\eta}\right), 
\end{align}
{\noindent where $u_\circ$ is the location of the serving BS, $\ell_\circ$ is the location of the harvesting device, $\zeta$ is the harvesting efficiency, $P$ is the constant transmit power for the BSs, ${g}_i$ is the channel gain between the harvesting device and the $i$-th BSs. Different from the depleted energy, the harvesting energy is a function of $r_n$ as well as the relative locations between the harvesting device and all other BSs. Exploiting the dominant contribution of the serving BS to the harvested energy, we resort to the following approximation:
{\begin{approximation} \label{app_EH}
		All devices within the same EH-class experience independent and identically distributed energy harvesting processes, which is obtained by fixing $r_n$ and averaging over all realizations of channel gains and non-serving BSs locations $\boldsymbol{\Psi} \setminus\{u_{\circ}\}$.
	\end{approximation}}}

{Exploiting Approximation~\ref{app_EH}, we characterize the  harvested energy ($F_{\mathcal{P}^{[n]}_H}$) in the next lemma. }

\begin{lemma}\label{lemma_harvested_power}
	The CDF of the harvested energy in a single-tier IoT network in a generic time slot is given by: 
		\begin{align}\label{energy_harvest}
		F_{\mathcal{P}_H^{[n]}}\!\!\left(x \right)\!=&\frac{1}{2}-\frac{1}{\pi}\!\! \int\limits_{0}^{\infty} \frac{1}{t} \text{Im} \left\{ \exp\left\{ -j \;t \;x \right\} \mathscr{L}_{\mathcal{P}_H^{[n]}}\left(-j\;t\right)\right\} dt,
		\end{align}
		where, $\mathscr{L}_{\mathcal{P}_H^{[n]}}$ is the Laplace transform of the harvested energy at each time slot, which can be evaluated by: 
		\begin{align}\label{harvest_LT}
			\mathscr{L}_{\mathcal{P}_H^{[n]}}\!(s)\!=&\frac{\exp\left\{\!\!-2\pi \lambda s T_s\zeta P \; {r}_n^{2-\eta}  \frac{ {}_2F_1\left(\!1,1-\frac{2}{\eta},2-\frac{2}{\eta},-s T_s\zeta P {r}_n^{-\eta} \right)}{\eta -2} \!\!\right\} }{1+s T_s\zeta P {r}_n^{-\eta}}\notag \\
			 \overset{(\eta=4)} =& \frac{\exp\left\{-\pi \lambda \sqrt{s T_s\zeta P} \; \arctan\left(\sqrt{s T_s\zeta P {r}_n^{-4}} \right) \right\} }{1+s T_s\zeta P {r}_n^{-4}}.
		\end{align}
	\begin{proof}
	By following \cite{hesham_tutorial}, the LT in \eqref{harvest_LT} is obtained by using the probability generating function (PGFL) of the PPP \cite{ martin_book}, the independence of the PPP in different regions \cite{ martin_book}, and the LT for the exponential distribution of $ {g}_\circ$. By utilizing Gil-Pelaez theorem \cite{Gill},  \eqref{energy_harvest}  is obtained.
	\end{proof}
\end{lemma}

{It is worth noting that the Laplace transform in \eqref{harvest_LT} is a function of $r_{n}$, and hence, accounts for the location-dependent EH-classes. Let $\mathbf{p}^{[n]}=[p^{[n]}_0,p^{[n]}_1,\dots, p^{[n]}_l, \dots, p^{[n]}_L]$, where $p^{[n]}_l$ is the probability that a device in EH-class $n$ harvests $l$ energy units in one time slot. Then, by rounding  the amount of the harvested energy down to the nearest integer, the probability of harvesting $0\leq l \leq L$ units of energy is}
\begin{align}\label{energy_levels} 
p^{[n]}_l= & \begin{cases} 
F_{\mathcal{P}_H^{[n]}}\left((l+1)w \right)-F_{\mathcal{P}^{[n]}_H}\left(lw \right),  & \text{if} \quad 0 \leq l < L, \\
1-	F_{\mathcal{P}^{[n]}_H}\left(Lw \right), & \text{if} \quad l=L.
\end{cases}
\end{align}

\subsubsection{Packet Transmission} {Consider a randomly selected BS located at an arbitrary location $u_\circ \in \mathbb{R}^2$ and denote the Voronoi cell of that test BS by $\mathcal{V}_\circ$. The transmission success probability of a randomly selected device within $\mathcal{V}_\circ$ can be expressed as
{	\begin{align} \label{SINR_RA}
	p_c&=\mathbb{P}\left\{ \underset{\mathcal{E}_\RN{1}}{\underbrace{ \frac{\rho h_\circ}{\sigma^2 + \mathcal{I}_{\text{Intra} }+\mathcal{I}_{\text{Inter}}}>\theta}},  \underset{\mathcal{E}_\RN{2}}{\underbrace{ h_\circ> h_i \; \forall h_i \in \bold{h}_{\mathcal{V}_\circ} \setminus h_\circ }} \mid  \forall h_i \in \bold{h}_{\mathcal{V}_\circ}  >\tau  \right\}, \notag \\
	&=\mathbb{P}\left\{ \frac{\rho h_\circ}{\sigma^2 + \mathcal{I}_{\text{Intra}}+\mathcal{I}_{\text{Inter}}}>\theta \;\mid  h_\circ> h_i \;  \forall h_i \in \bold{h}_{\mathcal{V}_\circ}  \setminus h_\circ , \forall h_i \in \bold{h}_{\mathcal{V}_\circ}  >\tau \right\} \notag \\ & \quad \quad \quad \quad \quad \quad \quad \quad \times \mathbb{P}\left\{h_\circ> h_i \;  \forall h_i \in \bold{h}_{\mathcal{V}_\circ}  \setminus h_\circ \right\}.
	\end{align}}
\noindent{where} $h_\circ$ is the intended device channel gain, $h_i$ is the intra-cell interfering devices channel gains,  $\bold{h}_{\mathcal{V}_\circ} $ is the vector of channel gains for all Tx-eligible and $P_T$-capable devices within $\mathcal{V}_\circ$. Also,  $\mathcal{I}_{\text{Intra}}$ and $	\mathcal{I}_{\text{Inter}}$ are, respectively, the intra-cell and the inter-cell interference from all Tx-eligible and $P_T$-capable devices, which are given by 
\begin{subequations}
	\begin{align}
\mathcal{I}_{\text{Intra}} &= \sum\limits_{\ell_i \in \bf{\Phi} \cap \mathcal{V}_\circ } \mathbbm{1}_{\{h_i\geq \tau \;\cap\; m_i\geq 1 \;\cap\; l_i\geq d^{[n]}\}}  \rho \ \mathtt{h}_i\label{eq:intra}\\
\mathcal{I}_{\text{Inter}} &=  \sum\limits_{\ell_k \in \bf{\Phi} \setminus \mathcal{V}_\circ }  \mathbbm{1}_{\{h_k\geq \tau \;\cap\; m_k\geq 1 \;\cap\; l_k\geq d^{[n]}\}} \mathtt{P}_{k} \mathtt{h}_k \left\|u_\circ-\ell_k\right\|^{-\eta} \label{eq:inter}
	\end{align}
	\label{eq:inter_intra}
\end{subequations}
\noindent{where} $\mathtt{P}_{k}$ and $\ell_k$ are, receptively, the transmission power and the  location of the $k$-th inter-cell interfering device. The three conditions within the indicator functions in \eqref{eq:inter_intra} ensure that the interfering devices have non-empty buffers $m_{\{\cdot\}} \geq 1$, Tx-eligible $h_{\{\cdot\}} \geq \tau$, and $P_T$-capable $l_{\{\cdot\}} \geq d^{[n]}$.  It is worth mentioning that the two highlighted events  $\mathcal{E}_\RN{1}$  and $\mathcal{E}_\RN{2}$ in \eqref{SINR_RA} jointly describe the SINR capture model with several conflicting intra-cell transmissions within $\mathcal{V}_\circ$. The event  $\mathcal{E}_\RN{2}$ dictates that the test BS can only decode the dominating signal within $\mathcal{V}_\circ$. Due to the employed path-loss inversion power control, the event $\mathcal{E}_\RN{2}$ reduces to the highest channel gain. The event $\mathcal{E}_\RN{1}$ ensures that the dominating signal power satisfies the required SINR threshold $\theta$. }

The inter-cell interference expression in \eqref{eq:inter} imposes two challenges for a tractable exact characterization of the success probability in \eqref{SINR_RA}. First, the transmission  powers $\mathtt{P}_{k}$ of adjacent devices are correlated, which is due to path-loss inversion power control along with the correlated devices distances~\cite{elsawy2014stochastic, marco_uplink, uplink2_jeff, 6516885}. Second, the inter-cell interference is a function of the relative locations of the interfering devices with respect to the test BS. To maintain the tractability of the analysis, we resort to the following two approximations 

{\begin{approximation} \label{app2}
		The correlations between the transmission powers of adjacent devices are ignored.
	\end{approximation}
	\begin{remark} \label{rem2}
		Approximation \ref{app2} assumes that the devices invert their  path-loss to the serving BS according to i.i.d transmit powers. Such approximation is widely utilized to maintain the mathematical tractability in the literature  \cite{uplink_alamouri,Meta_Haenggi_Control, Meta_Elsawy, Gharbieh_tcom,elsawy2014stochastic, marco_uplink, uplink2_jeff, 6516885}. It is worth to mention that the Monte Carlo simulations in Section~\ref{Results} accounts for the spatial correlations, and hence, the results validate our model.
	\end{remark}
	
	\begin{approximation} \label{app1}
		 The successful transmission probability at a typical BS is used to approximate the successful transmission probabilities of all devices in the network. This implies that such probabilities are location-independent and uncorrelated over different time slots.
	\end{approximation}
	\begin{remark} \label{rem1}
		The success probability in \eqref{SINR_RA} is dominated by the intended signal power and intra-cell interference, which are independent of the relative locations between the serving BS, intended device, and interfering device. Hence, the SINR is dominated by the realizations of the channel gains, which are independent across different devices and time slots. Hence, the success probability at a typical BS obtained via spatial averaging is a good representation for the success probabilities of all devices in the network, which is in compliance with findings in~\cite{Meta_Haenggi_Control, Meta_Elsawy, Gharbieh_tcom}. 
	\end{remark} 
{Let $\delta=\mathbb{E}\{\mathbbm{1}_{\{m_i\geq 1 \;\cap\; l_i\geq d^{[n]}\}}\}$ be the probability that a generic device in the network has non-empty buffer and is $P_T$-capable. The analysis to characterize $\delta$ is presented in the queueing theory part in Section~\ref{sic:Performance_Analysis2}. Exploiting Approximations 2 \& 3 and accounting for the mean-field effect of the aggregate interference, the success probability in \eqref{SINR_RA} is characterized in the following lemma}

\begin{lemma} \label{lem:out_NB_IoT}
	The transmission success probability in the depicted PPP network with opportunistic GF-UL, where only the packet that has the highest SINR is received correctly is given by	
	
		\small
			\begin{align}\label{eq:TX_NB_success}
			p_c\!=\! \mathbb{E}_{\mathcal{N}}\! &\left\{\left\{\!\sum\limits_{k=1 }^{\mathcal{N}+1} \binom{\mathcal{N}+1}{k}(-1)^{k+1}\left[  F_{{\mathcal{I}_{\text{Inter}}}}\left(\frac{\tau \rho}{\theta}-\sigma^2\right)F_{{\mathcal{I}_{\rm Intra}}}\left(\frac{\tau \rho}{\theta}-\sigma^2 \big| {\mathcal{N}= \mathtt{n}}\right) \right.\right. \right.\notag \\ 
			&+ \mathscr{L}_{\mathcal{I}_{\text{Inter}}}\left(\frac{k \; \theta}{\rho} \right) \mathscr{L}_{\mathcal{I}_{\rm Intra}} \left(\frac{k \; \theta}{\rho}\big| {\mathcal{N}=\mathtt{n}} \right) \bar{F}_{{\mathcal{I}_{\text{Inter}}}}\left(\frac{\tau \rho}{\theta}-\sigma^2\right)\bar{F}_{{\mathcal{I}_{\rm Intra}}}\left(\frac{\tau \rho}{\theta}-\sigma^2\big| {\mathcal{N}= \mathtt{n}}\right) \notag \\
			& + \mathscr{L}_{\mathcal{I}_{\text{Inter} }}\left(\frac{k \; \theta}{\rho} \right) \bar{F}_{{\mathcal{I}_{\text{Inter} }}}\left(\frac{\tau \rho}{\theta}-\sigma^2\right)F_{{\mathcal{I}_{\rm Intra}}}\left(\frac{\tau \rho}{\theta}-\sigma^2\big| {\mathcal{N}= \mathtt{n}}\right) \notag \\
			& +  \left. \left. \left. \mathscr{L}_{\mathcal{I}_{\rm Intra}} \left(\frac{k \; \theta}{\rho} \big| {\mathcal{N}=\mathtt{n}}\right) F_{{\mathcal{I}_{\text{Inter} }}}\left(\frac{\tau \rho}{\theta}-\sigma^2\right)\bar{F}_{{\mathcal{I}_{\rm Intra}}}\left(\frac{\tau \rho}{\theta}-\sigma^2\big| {\mathcal{N}=\mathtt{n}}\right)\right]\right\}/\left\{\mathcal{N}+1\right\}\right\},
			\end{align}
		\normalsize
	with
	\small
	{\begin{align}
		& \mathscr{L}_{\mathcal{I}_{\text{Inter} }}\!\left(\!\frac{k \; \theta}{\rho}\!\right) \!\approx \exp\left\{\!\!\! - \! 2 \;\frac{ \delta \;\Omega \mu^{\prime}}{\lambda} \;\frac{ \gamma\left(2,\pi \lambda \left(\frac{P_{max}}{\rho}\right)^\frac{2}{\eta} \right)}{ \left(1 - e^{-\pi \lambda \left(\frac{P_{max}}{\rho}\right)^{\frac{2}{\eta}}}\right)} (k\theta)^{2/\eta} \! \! \! \int\limits_{(k\theta)^{\frac{-1}{\eta}}}^{\infty}\! \! \!\left(\!1\! - \! \frac{\exp\{\!-\tau y^{-\eta }\!\}}{y^{-\eta} +1}\!\right)y \; dy \right\}, \label{Eq:laplase_Inter11}\\
		& \mathscr{L}_{\mathcal{I}_{\rm Intra}}\!\left(\!\frac{k \; \theta}{\rho} \big|{\mathcal{N}= \mathtt{n}} \right)\!\!= \left[\exp\!\left(\!- k \theta \tau \!\right)\!\frac{{n}+1}{1+k \; \theta} \left(\!\frac{1}{\mathtt{n}}\! -\!\frac{\Gamma(\mathtt{n}) \;\Gamma(2+k \; \theta)}{\Gamma(2+\mathtt{n}+k \; \theta)} \right)\right]^n,
		\end{align}}
		\normalsize
	where  $\mu^{\prime}=\mu/n_{\text{\rm c}}$ and  \eqref{Eq:laplase_Inter11} is not exact due to ignoring the spatial correlations among the transmission powers of the devices.  The expectation $\mathbb{E}_{\mathcal{N}}\{\cdot\}$ is with respect to the distribution of the number of intra-cell interferers $\mathcal{N}$, which is given by: 
		\begin{align}\label{pdf_users1}
		\mathbb{P}\{\mathcal{N} = {n}\} \approx \frac{\Gamma({n}+c)}{\Gamma({n}+1)\Gamma(c)} \frac{(\delta  \;\Omega \mu^{\prime})^{{n}} (\lambda c)^c}{(\delta   \;\Omega \mu^{\prime}+\lambda c)^{{n}+c}},
		\end{align}
		where $c=3.575$ is a constant related to the approximate PDF of the PPP Voronoi cell area in $\mathbb{R}^2$ {\normalfont \cite{6576413}}. $F_{{\mathcal{I}_{\text{Inter} }}}\!\!\left(x \right)$ is the CDF of the aggregated inter-cell interference which has the form of \eqref{eq:CDF_I} as:
	\small
{\begin{align}\label{eq:CDF_I}
\!F_{{\mathcal{I}_{\text{Inter} }}}\!\!\left(x \right)= &\frac{1}{2}\!-\!\frac{1}{\pi}\!\! \int\limits_{0}^{\infty} \! \frac{1}{t} \text{Im} \left\{\!\!\ \!\!\!\ \! \exp\left\{ -j \;t \;x  \right\}{\exp\left\{\!\!\ \! - \! 2 \; \frac{ \delta \;\Omega \mu^{\prime}}{\lambda} \frac{ \gamma\left(2,\pi \lambda \left(\frac{P_{max}}{\rho}\right)^\frac{2}{\eta} \right)}{1 - e^{-\pi \lambda \left(\frac{P_{max}}{\rho}\right)^{\frac{2}{\eta}}}}\; \! \!\rho^{2/\eta} \! \! \! \! \! \int\limits_{\rho^{\frac{-1}{\eta}}}^{\infty}\! \! \! \!\left(\!1\! - \! \frac{\exp\{j \; \tau \;t \;z^{-\eta }\}}{-j\;t\;z^{-\eta} +1}\!\right)z \; dz\!\!\ \! \right\}}\right\} dt. 
\end{align}}
\normalsize
	
$F_{{\mathcal{I}_{\rm Intra}}}\!\!\left(x \big| \mathcal{N}= \mathtt{n}\right)$ is the conditional CDF of the aggregated intra-cell interference which is approximated via the gamma distribution as:
\begin{align}\label{eq:CDF_In}
F_{{\mathcal{I}_{\rm Intra}}}\!\!\left(x \big| \mathcal{N}= \mathtt{n}\right) \!\!\left(x \right)\approx &\frac{1}{\Gamma(\alpha)}\gamma(\alpha,\beta x),
\end{align}
with a shape parameter $\alpha$,
\begin{align}\label{eq:shape_par}
\alpha= \frac{\left(n\tau+ n+1 -  H_{n+1}\right)^2}{-2 \nu +n+1+\pi^2/6-2\psi^{(0)}(n+2)-\psi^{(1)}(n+2)},
\end{align}
and a rate parameter $\beta$,
\begin{align}\label{eq:rate_par}
\beta= \frac{n\tau+n+1 -  H_{n+1}}{\rho\left(-2\nu +n+1+\pi^2/6-2\psi^{(0)}(n+2)-\psi^{(1)}(n+2)\right) }.
\end{align}

\begin{proof}
	See Appendix \ref{proof1}.
\end{proof}
	
\end{lemma}
{For the special case of $\tau=0$, in which all $P_T$-capable devices are also Tx-eligible, the probability of successful transmission reduces to: 
\small
\begin{align}\label{eq:TX_NB_success11}
p_c\!=\! \mathbb{E}_{\mathcal{N}}\! &\left\{\left\{\!\sum\limits_{k=1 }^{\mathcal{N}+1} \binom{\mathcal{N}+1}{k}(-1)^{k+1}\left[ \exp\!\left(\!- \frac{k \theta \sigma^2}{\rho} \!\right) \mathscr{L}_{\mathcal{I}_{\text{Inter} }}\left(\frac{k \; \theta}{\rho} \right) \mathscr{L}_{\mathcal{I}_{\rm Intra}} \left(\frac{k \; \theta}{\rho}\big|{\mathcal{N}=\mathtt{n}} \right) \right]\right\}/\left\{\mathcal{N}+1\right\}\right\},
\end{align}
\normalsize
with
\small
\begin{align}
& \mathscr{L}_{\mathcal{I}_{\text{Inter} }}\!\left(\!\frac{k \; \theta}{\rho}\!\right) \!\approx \exp\left\{\!\!\! -2 \; k \; \theta\; \frac{\delta \; \mu^{\prime}}{\lambda} \; \frac{{}_2F_1\left(1,1-\frac{2}{\eta},2-\frac{2}{\eta},-k\;\theta\right)}{\eta-2} \right\}, \label{Eq:laplase_Inter22}\\
& \mathscr{L}_{\mathcal{I}_{\rm Intra}}\!\left(\!\frac{k \; \theta}{\rho} \big| {\mathcal{N}= \mathtt{n}} \right)\!\!= \left[\frac{{n}+1}{1+k \; \theta} \left(\!\frac{1}{\mathtt{n}}\! -\!\frac{\Gamma(\mathtt{n}) \;\Gamma(2+k \; \theta)}{\Gamma(2+\mathtt{n}+k \; \theta)} \right)\right]^n.
\end{align}
\normalsize }
\subsection{Queueing Theory Analysis} \label{sic:Performance_Analysis2}
{This section develops queuing theory analysis to find the joint distribution of the data buffer and battery states. As shown in Fig.~\ref{fig:MC_buf} only a single unit-step transition can occur between buffer states in each time slot. Hence, the data buffer states exhibit a quasi-birth-death (QBD) behavior. Utilizing the matrix analytic method for QBDs, the 2D DTMC of the joint data buffer and battery states for a device in EH-class $n$ can be constructed as follows}
\small
\begin{align}
\mathbf{P}^{[n]}\!\!=\!\!\!\begin{bmatrix}
\bar{a} \mathbf{H}^{[n]} \!&\! a \mathbf{H}^{[n]} \!&\!  \!&\!  \!\!&\!\!   \\
\Omega\bar{a} \mathbf{S}^{[n]} \!&\!\Omega\!\left[ a \mathbf{S}^{[n]} \!+\!\bar{a} \mathbf{F}^{[n]}\right] \!+\!\bar{\Omega} \bar{a} \mathbf{H}^{[n]} \!&\! \Omega a \mathbf{F}^{[n]}\! +\! \bar{\Omega}a \mathbf{H}^{[n]}\!&\!  \!\!&\!\!  \\
\!&\!\Omega\bar{a} \mathbf{S}^{[n]} \! &\! \Omega\!\left[ a \mathbf{S}^{[n]}\! +\!\bar{a} \mathbf{F}^{[n]}\right]\! +\!\bar{\Omega} \bar{a} \mathbf{H}^{[n]} \!&\! \Omega a \mathbf{F}^{[n]}\! + \!\bar{\Omega}a \mathbf{H}^{[n]} \!\!&\!\!   \\
\!& \! \!& \!    \!&\!\ddots  \!\!&\!\!\ddots  \\
\!& \! \!& \!  \!&\! \Omega\bar{a} \mathbf{S}^{[n]} \!\!&\!\! \Omega \! \left[1\!-\! \bar{a} \mathbf{S}^{[n]}\right]\!+ \!\bar{\Omega}\mathbf{H}^{[n]}  
\end{bmatrix},
\label{NB_MAT}
\end{align}
\normalsize
{where the matrix $\mathbf{P}^{[n]}$ shows the unit-step transitions of the data packets and the sub-matrices $\mathbf{H}^{[n]}$, $\mathbf{F}^{[n]}$, and $\mathbf{S}^{[n]}$ capture all the underlying battery states transition probabilities during, respectively, energy harvesting, transmission failure, and transmission success. Note that $\mathbf{H}^{[n]}$ represents the battery transitions when the device is not Tx-eligible, and hence, $\mathbf{H}^{[n]}$ has only energy harvesting transitions and does not contribute to packet departures. On the other hand, $\mathbf{S}^{[n]}$ represents battery transitions upon transmission success, and hence,  $\mathbf{S}^{[n]}$ compasses only energy depletion transitions and contributes to packet departures. The matrix $\mathbf{F}^{[n]}$ takes effect when the device is Tx-eligible but does not contribute to packet departures, which implies either i) energy harvesting due to insufficient battery units or ii) energy depletion with transmission failure. Hence, the matrix  $\mathbf{F}^{[n]}$ compasses both the energy harvesting and depletion transitions.}

{The matrix $\mathbf{P}^{[n]}$ has $(M+1)\times(M+1)$ states for the data buffer, where each state encompasses all $(L+1)\times(L+1)$ possible battery states transition. Hence, the  $\mathbf{P}^{[n]}$ is a square stochastic matrix of dimension $[(M+1).(L+1)]\times[(M+1).(L+1)]$ that captures all the possible joint buffer and battery states. The  sub-matrices $\mathbf{H}^{[n]}$, $\mathbf{F}^{[n]}$, and $\mathbf{S}^{[n]}$ are of size $(L+1)\times (L+1)$ and can be expressed as 
\small
\begin{align}
\mathbf{H}^{[n]}=\begin{bmatrix}
\mathbf{H}_{1}^{[n]} & \mathbf{H}_{1,2}^{[n]}  \\ 
\mathbf{0} & \mathbf{H}_{2}^{[n]}  \\ 
\end{bmatrix},
\quad \text{} \quad 
\mathbf{F}^{[n]}=\begin{bmatrix}
\mathbf{F}_{1}^{[n]} & \mathbf{F}_{1,2}^{[n]}  \\ 
\mathbf{F}_{2,1}^{[n]} & \mathbf{F}_{2}^{[n]}  \\ 
\end{bmatrix}, \quad \text{and} \quad 
\mathbf{S}^{[n]}=\begin{bmatrix}
\mathbf{0} & \mathbf{0}  \\ 
\mathbf{S}_{2,1}^{[n]} & \mathbf{S}_{2}^{[n]}  \\ 
\end{bmatrix}.
\label{sub_all}
\end{align}
\normalsize
Where 
\small
\begin{align}
\mathbf{H}_1^{[n]}= \mathbf{F}_1^{[n]} =\begin{bmatrix}
p_0 & p_1 &  \dots & {p_{d^{[n]}-2}} & p_{d^{[n]}-1}  \\
0&p_0 & \dots  & {p_{d^{[n]}-3}}  & p_{d^{[n]}-2}  \\
\vdots& \ddots &  \ddots    &\vdots  &\vdots  \\
0& \dots & 0  & p_0 & p_1  \\ 
0& \dots & 0  & 0 & p_0 \\ 
\end{bmatrix},
\label{sub_HF1}
\end{align}
\normalsize
and
\small
\begin{align}
\mathbf{H}_{1,2}^{[n]} = \mathbf{F}_{1,2}^{[n]}=\begin{bmatrix}
p_{d^{[n]}} &p_{d^{[n]}+1} &  \dots  & p_{L-1} & 1-\sum\limits_{l=0}^{L-1}p_l \\
p_{d^{[n]}-1}& p_{d^{[n]}-2} & \dots  & p_{L-2} & 1-\sum\limits_{l=0}^{L-2}p_l \\
\vdots& \ddots &  \ddots    &\vdots  &\vdots  \\
p_2& p_3 &\dots   & p_{L-d^{[n]}-2} & 1-\sum\limits_{l=0}^{L-d^{[n]}-2}p_l  \\ 
p_1& p_2 & \dots & p_{L-d^{[n]}-1} & 1-\sum\limits_{l=0}^{L-d^{[n]}-1}p_l \\ 
\end{bmatrix}.
\label{sub_HF2}
\end{align}
\normalsize
where $p^{[n]}_l$ is given in \eqref{energy_levels}. As long as the device is not Tx-eligible, it stays in the energy harvesting mode even if $P_T$-capable, and hence $\mathbf{H}_{2}^{[n]}$ has the following energy harvesting transitions
\small
\begin{align}
\mathbf{H}_{2}^{[n]}=\begin{bmatrix}
p_0 & p_1 &  \dots  & p_{L-\chi-1} & 1-\sum\limits_{l=0}^{L-\chi-1}p_l \\
0&p_0 & \dots  & p_{L-\chi-2} & 1-\sum\limits_{l=0}^	{L-\chi-2}p_l \\
\vdots& \ddots &  \ddots    &\vdots  &\vdots  \\
0& \dots & 0  & p_0 & 1-p_0  \\ 
0& \dots & 0  & 0 & 1 \\ 
\end{bmatrix},
\label{sub_H2}
\end{align}
\normalsize
where $\chi = \min(d^{[n]}-1, L-d^{[n]})$. When the device is Tx-eligible and $P_T$-capable, it has to attempt an uplink transmission, which can be successful or not. In either cases, each transmission attempt depletes $d^{[n]}$ energy units from the battery. The remaining battery levels after an uplink transmission may make the device $P_T$-incapable, which is captured by $\mathbf{S}_{2,1}^{[n]}$ and  $\mathbf{F}_{2,1}^{[n]}$ as 
\small
\begin{align}
\mathbf{S}_{2,1}^{[n]}= \left\{\begin{matrix}
 p_c \begin{bmatrix}
\mathbf{I}_{\chi} \\ 
\mathbf{0} 
\end{bmatrix}
 &  \text{if} \; L>2 d^{[n]}-1 \\ 
& \\
 p_c  \; \mathbf{I}_{\chi} 
 & \text{if} \; L=2 d^{[n]}-1 \\ 
& \\
 p_c \begin{bmatrix}
\mathbf{I}_{\chi} & 
\mathbf{0} 
\end{bmatrix} &  \text{if}\; L<2 d^{[n]}-1
\end{matrix}\right. \quad  \quad\text{and}\quad \quad 
\mathbf{F}_{2,1}^{[n]}= \left\{\begin{matrix}
\bar{p}_c \begin{bmatrix}
\mathbf{I}_{\chi} \\ 
\mathbf{0} 
\end{bmatrix}
&  \text{if} \; L>2 d^{[n]}-1 \\ 
& \\
\bar{p}_c  \; \mathbf{I}_{\chi} 
& \text{if} \; L=2 d^{[n]}-1 \\ 
& \\
\bar{p}_c \begin{bmatrix}
\mathbf{I}_{\chi} & 
\mathbf{0} 
\end{bmatrix} &  \text{if}\; L<2 d^{[n]}-1
\end{matrix}\right.
\label{sub_SF1}
\end{align}
\normalsize
If the device has enough energy units in its battery, it can transmit and remain $P_T$-capable, which is captured by $\mathbf{S}_{2}^{[n]}$ and  $\mathbf{F}_{2}^{[n]}$ as 
\small
\begin{align}
\mathbf{S}_{2}^{[n]}= \left\{\begin{matrix}
p_c \begin{bmatrix}
\mathbf{0} & \mathbf{0} \\
\mathbf{I}_{L-2 \chi+1} & \mathbf{0} 
\end{bmatrix}
&  \text{if} \; L>2 d^{[n]}-1 \\ 
& \\
\mathbf{0} 
& \text{otherwise}
 \end{matrix}\right.
 \quad \quad \text{and} \quad \quad
 \mathbf{F}_{2}^{[n]}= \left\{\begin{matrix}
 \bar{p}_c \begin{bmatrix}
 \mathbf{0} & \mathbf{0} \\
 \mathbf{I}_{L-2\chi+1} & \mathbf{0} 
 \end{bmatrix}
 &  \text{if} \; L>2 d^{[n]}-1 \\ 
 & \\
 \mathbf{0} 
 & \text{otherwise}
 \end{matrix}\right.
\label{sub_SF2}
\end{align}
\normalsize
Once $\mathbf{P}^{[n]}$ in \eqref{NB_MAT} is constructed, the steady state distribution of all buffer and battery state can be obtained. Let $\mathbf{x}^{[n]}=[\mathbf{x}^{[n]}_0, \mathbf{x}^{[n]}_1, \mathbf{x}^{[n]}_2,\ \hdots, \mathbf{x}^{[n]}_M\ ]$ be the stationary distribution for the $n^{th}$ class, where $\mathbf{x}^{[n]}_m=[x^{[n]}_{m,1}, x^{[n]}_{m,2}, \dots, x^{[n]}_{m,L}]$ represents the probability of having $m$ packets in the data buffer and  $x^{[n]}_{m,l}$ is the probability of having $m$ packets and $l$ energy levels. The steady-state solution $\mathbf{x}^{[n]}$ for the queuing model is obtained by solving the following system of equations
\begin{align} \label{hhss}
\mathbf{x}^{[n]}  \mathbf{P}^{[n]}=\mathbf{x}^{[n]} \quad  \text{and}  \quad\mathbf{x}^{[n]}  \mathbf{e}&=1,
\end{align}
where $\mathbf{e}$ is a ones column vector of size $\left((L+1)(M+1)\right)$.}

{The system of linear equations in \eqref{hhss} can be easily solved if all parameters in $\mathbf{P}^{[n]}$ are known. However, $\mathbf{P}^{[n]}$ requires $p_c$ as shown in \eqref{sub_SF1} and \eqref{sub_SF2}. The transmission probability $p_c$ is characterized in Lemma~\ref{lem:out_NB_IoT} as a function of  $\delta$, which is the probability that a generic device has non-empty buffer and is $P_T$-capable. The probability $\delta$ can be expressed in terms of steady state distribution $\mathbf{x}^{[n]}$ for all EH-classes as 
\begin{align} \label{delta}
\delta=\sum\limits_{n =1}^{N} \frac{1}{N}\sum\limits_{m=1 }^{M} \;\sum\limits_{l=d^{[n]}}^{L} x^{[n]}_{m,l}.
\end{align}
Hence, $p_c$ in \eqref{eq:TX_NB_success} and $\mathbf{x}^{[n]}$ in \eqref{hhss} are interdependent. While no explicit closed form solution can be obtained for either $p_c$ or $\mathbf{x}^{[n]}$, both can be obtained via the  iterative solution presented in the next section. } 
\subsection{Iterative Solution \& Performance Assessment} \label{sic:Performance_Analysis3}
\begin{figure}
	\begin{algorithm}[H]
		\small
		Initialize   $i=1$ and $x^{[n]}_{\circ, \circ}=1 \; \forall n$.\\
		\While { $\left|\mathbf{x}_{[i]} - \mathbf{x}_{[i-1]} \right| \geq \epsilon $  } {
			1- Calculate $\delta^{[n]}_{[i]} \; \forall n$ using $\mathbf{x}^{[n]}_{[i-1]}$ in \eqref{delta}.\\
			2- Evaluate $p_c$ in  \eqref{eq:TX_NB_success}.\\
			3- Construct $\mathbf{H}^{[n]}_{[i]}$, $\mathbf{S}^{[n]}_{[i]}$, and $\mathbf{F}^{[n]}_{[i]} \;\forall n$  using $p_c$ as in \eqref{sub_all}, respectively. \\
			4- Construct $\mathbf{P}^{[n]}_{[i]} \; \forall n$ as in \eqref{NB_MAT}.\\
			5- Solve the systems of equation in \eqref{hhss} to get $\mathbf{x}^{[n]}_{[i]} \; \forall n$.\\
			6- Increment $i$.}
		Return $\mathbf{x} \leftarrow \mathbf{x}_{[i]}$ and $p_c$.		
		\caption{Iterative Solution for the Steady-State Distribution $\mathbf{x}$.}
		\label{dist_algo}
		\normalsize
	\end{algorithm}
\end{figure}

{ Based on the analytical expression in Section~\ref{sic:Performance_Analysis1} and Section \ref{sic:Performance_Analysis2},  it is clear that the queuing theory and the  stochastic geometry analysis are interdependent. The interference LT, and hence, the successful packet departure probability $p_c$  at one device depends on the mode of operation and transmission powers of all other devices in the network. As such, to obtain $p_c$ for a given device, the state probabilities $\mathbf{x}$ of the DTMC for all other devices is required. The other face of the coin is also true, $p_c$ is required in Section~\ref{sic:Performance_Analysis2} for queuing analysis to solve the DTMC and obtain the underlying state probabilities for all devices. Such an interdependence can be solved iteratively as in Algorithm~1, which converges by virtue of the fixed point theorem~\cite{Heanggi_Ra1, Zhuang, interacting_queues2, SarElaFouNou:07, Gharbieh_tcom}.\footnote{It is important to note that Algorithm~1 is conducted offline to characterize the network performance and come-up with long-term network design (e.g., $\tau$ and $\rho$). Such network design do not need to be changed as long as the underlying network statistical parameters (e.g, $\lambda$, $\mu$, $n_c$, $\zeta$, $\eta$, and $a$) remain fixed.} The iterative algorithm is equivalent to solve the system for each class
\begin{align} 
\mathbf{x}^{[n]}_{[j]}  \mathbf{P}^{[n]}(\mathbf{x}^{[n]}_{[j-1]} )=\mathbf{x}^{[n]}_{[j]} \quad  \text{and} \quad\mathbf{x}^{[n]}_{[j]}  \mathbf{e}&=1,
\end{align}
where $\mathbf{x}^{[n]}_{[j]}$ is the state probabilities at the $j^{th}$ for the class $n$, $\mathbf{P}^{[n]}(\mathbf{x}^{[n]}_{[j-1]} )$ is updates based on \eqref{NB_MAT}, and $\mathbf{e}$ is a ones column vector of size $\left((L+1)(M+1)\right)$.}} After computing the steady-state distribution $\mathbf{x}$ and packet departure rate in Algorithm~1, a variety of performance metrics can be defined and evaluated as follows:
\begin{itemize}
	\item \textbf{Average buffer Size $\mathbb{E}\left\{\mathcal{Q}_{\text{\rm L}}^{[n]}\right\}$:}  Let $\mathcal{Q}_{\text{\rm L}}^{[n]}$ be the instantaneous buffer size of the test device in the $n^{th}$ class, then the average buffer size is given by
	\begin{align} \label{ave_queue}
	\mathbb{E}\left\{\mathcal{Q}_{\text{\rm L}}^{[n]}\right\} &=\!\! \sum_{m=1}^{M} m   \;\mathbb{P} \left\{\mathcal{Q}_{\text{\rm L}}^{[n]}=m \right\} \!=\!\sum_{m=1}^{M} m \sum_{l=0}^{L} x^{[n]}_{m,l}.
	\end{align}
	
	\item \textbf{Average Packets Throughput $\mathbb{E}\left\{\mathcal{T}^{[n]}\right\}$:}  Let $\mathcal{T}^{[n]}$ be the instantaneous packet throughput for a test device in the $n^{th}$ class, which is the  successful transmitted packets per time slot, then the average packet throughput is given by
	\begin{align} \label{ave_throughput}
	\mathbb{E}\left\{\mathcal{T}^{[n]}\right\} &= \Omega \; p_c \; \delta^{[n]}.
	\end{align}

	\item \textbf{Average {Delay} $\mathbb{E}\left\{\mathcal{W}^{[n]}\right\}$}: Let $\mathcal{W}^{[n]}$ be how many time slots a given packet spent in the buffer of a test device in the $n^{th}$ class until successful transmission, then the average delay (i.e., overall the packets) can be evaluated by exploiting  Little's Law \cite{Little2008} as
	\begin{equation} \label{wait_gen}
	\mathbb{E}\left\{\mathcal{W}^{[n]}\right\}= \frac{\mathbb{E}\left\{\mathcal{Q}_{\text{\rm L}}^{[n]}\right\}}{\mathbb{E}\left\{\mathcal{T}^{[n]}\right\}}.
	\end{equation}

	\item \textbf{Packet Loss Probability ($L^{[n]}$)}: Due to the constraint of the limited buffer size, the packets are lost if they arrive when the buffer is full. As such, the packet loss probability for a test device in the $n^{th}$ class can be computed by 
		\begin{equation} \label{Loss}
	L^{[n]}= 1- \frac{\mathbb{E}\left\{\mathcal{T}^{[n]}\right\}}{a}. 
	\end{equation}
\end{itemize}

\section{Numerical Results} \label{Results}



\begin{figure}[t!]
	\centering
	\begin{minipage}{.45\textwidth}
		\centering
		\includegraphics[width=2.9 in ]{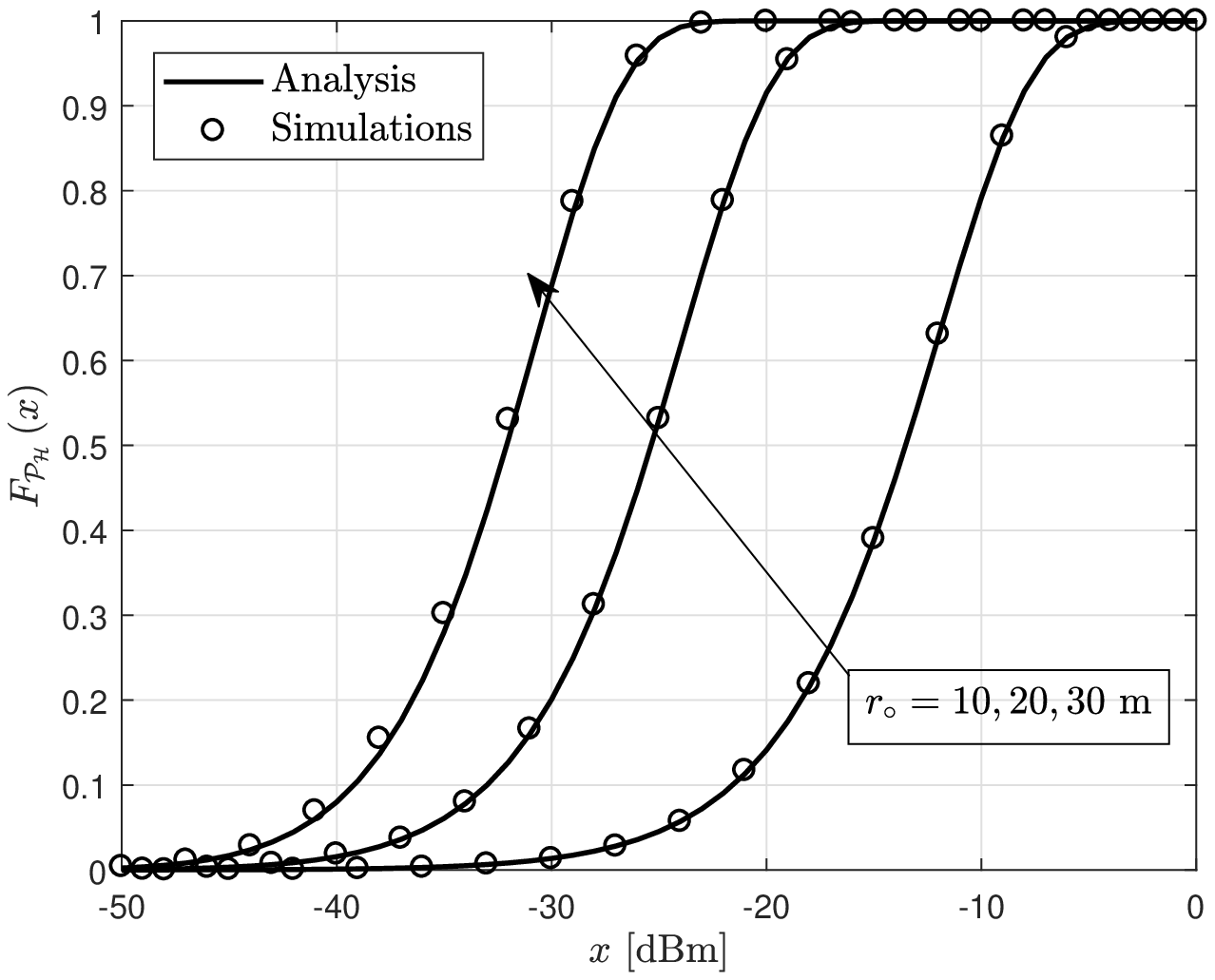}
		\captionof{figure}{$\; \;$ Verification of the harvested energy CDF in Lemma 1 $F_{\mathcal{P}^{[\circ]}_H}\left(x \right)$.}\label{harvest_figure}
	\end{minipage}%
	\hspace{10mm}
	\begin{minipage}{.45\textwidth}
		\centering
		\includegraphics[width=2.9 in ]{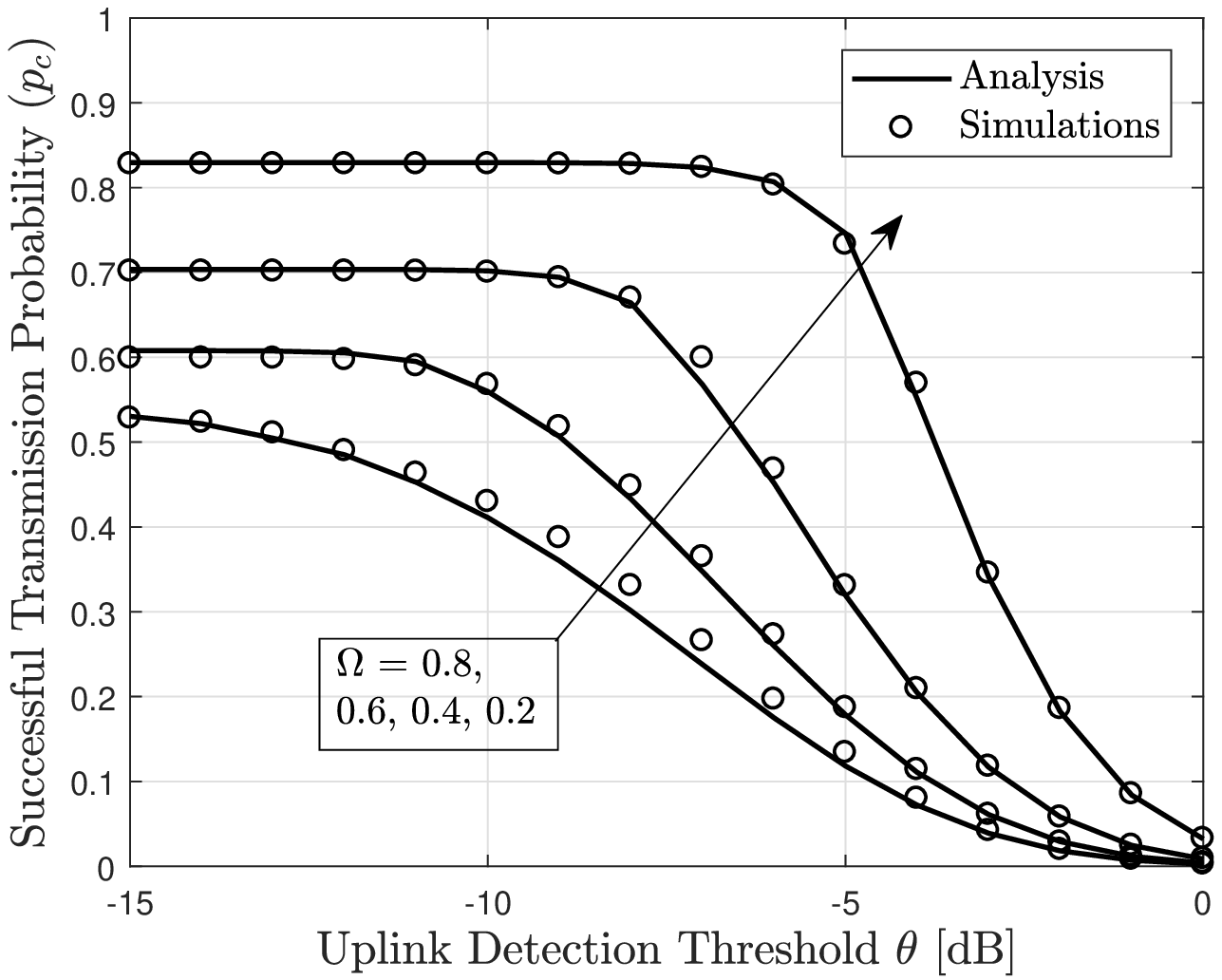}
		\captionof{figure}{$\;$Verification of the transmission success probability $p_c$ presented in Lemma 2.}\label{success_figure}
	\end{minipage}
\end{figure}

At first, we compare the proposed stochastic geometry analysis for the harvested energy with independent system-level simulations in Fig.~\ref{harvest_figure}, where the energy harvesting was not discretized, but rather, evaluated independently in each simulation run. For every simulation iteration, the BSs are distributed over a 100 km$^2$ via a PPP with $\lambda=35$ BS/km$^2$, $\eta=4$, $T_s=1$ ms, $\zeta P=28\;$dBm. The data are taken for a test device located at a distance $r_\circ= 10, 20,$ and $ 30 $ meters from the test BS. As can be seen, the analytical results and the simulation results match closely, which validate the proposed stochastic geometry analysis. Then, we compare the proposed stochastic geometry analysis for the probability of successful transmission with independent system level simulations in Fig.~\ref{success_figure}. {We choose $\lambda=35$ BS/km$^2$,  $\delta\mu= 70$ device/km$^2$/resource block, path-loss exponent $\eta =4$, noise power $\sigma= -120$ dBm, power control threshold $\rho =-126$ dBm, channel transmission probability $\Omega= 0.2, 0.4, 0.6, 0.8$, $\theta\!=\! [-15,0]$ dB.}  In each simulation run, the BSs are realized over a 100 km$^2$ via a PPP. The collected statistics are for test devices located within 1 km$^2$ from the origin. The close match between the analysis and simulation results validates the developed stochastic geometry analysis. It is worth noting that due to the spatiotemporal analysis of the model, the analytical performance assessment complexity is marginal when compared to that of the system-level simulations.



\begin{figure}[t!]
	\centering
	\begin{minipage}{.45\textwidth}
		\centering
		\includegraphics[width=2.9 in ]{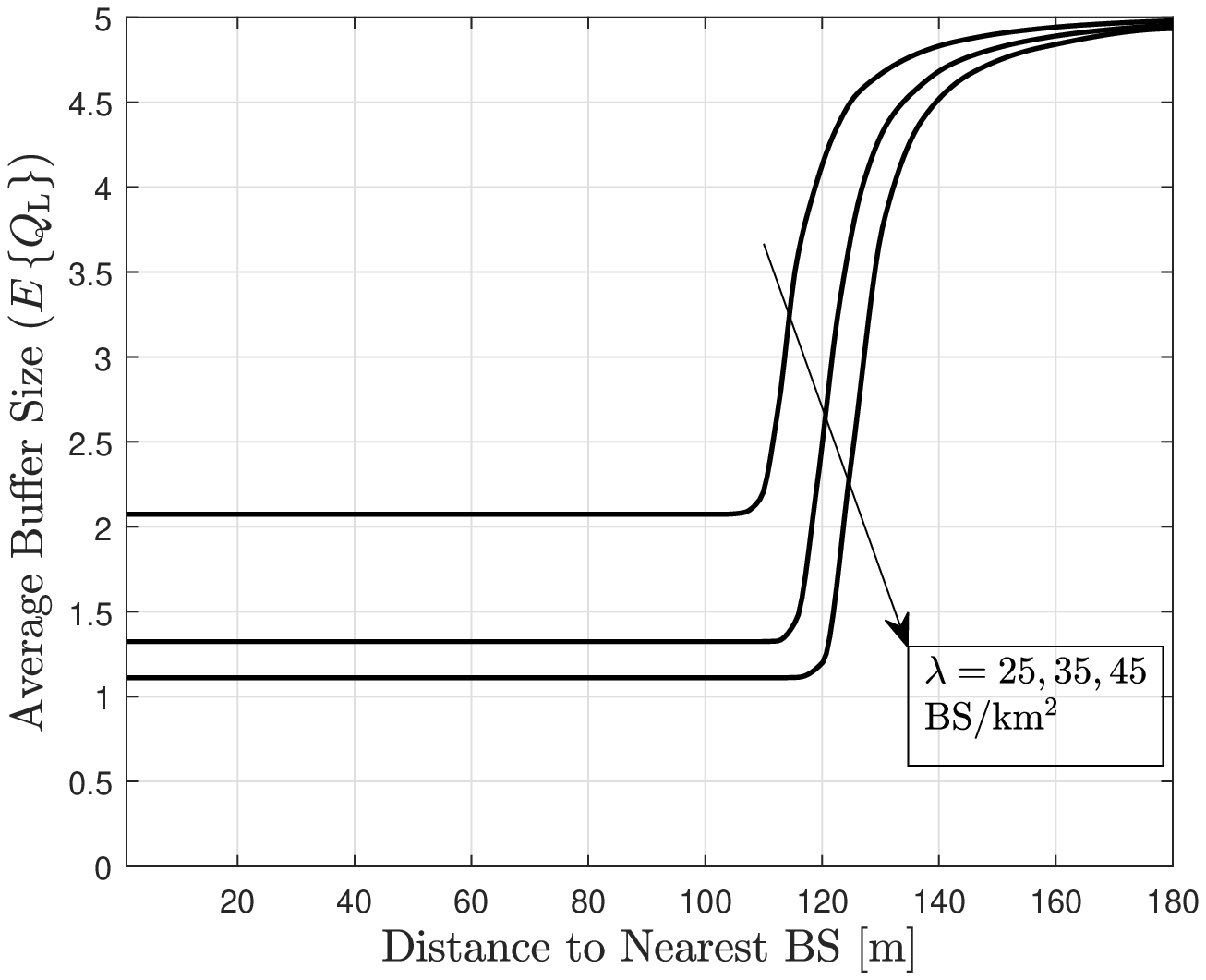}
		\captionof{figure}{$\;\;$The steady-state average buffer size as a function of the distance to the nearest BS in meters.}
		\label{fig_queue}
	\end{minipage}%
	\hspace{10mm}
	\begin{minipage}{.45\textwidth}
		\centering
		\includegraphics[width=2.9 in ]{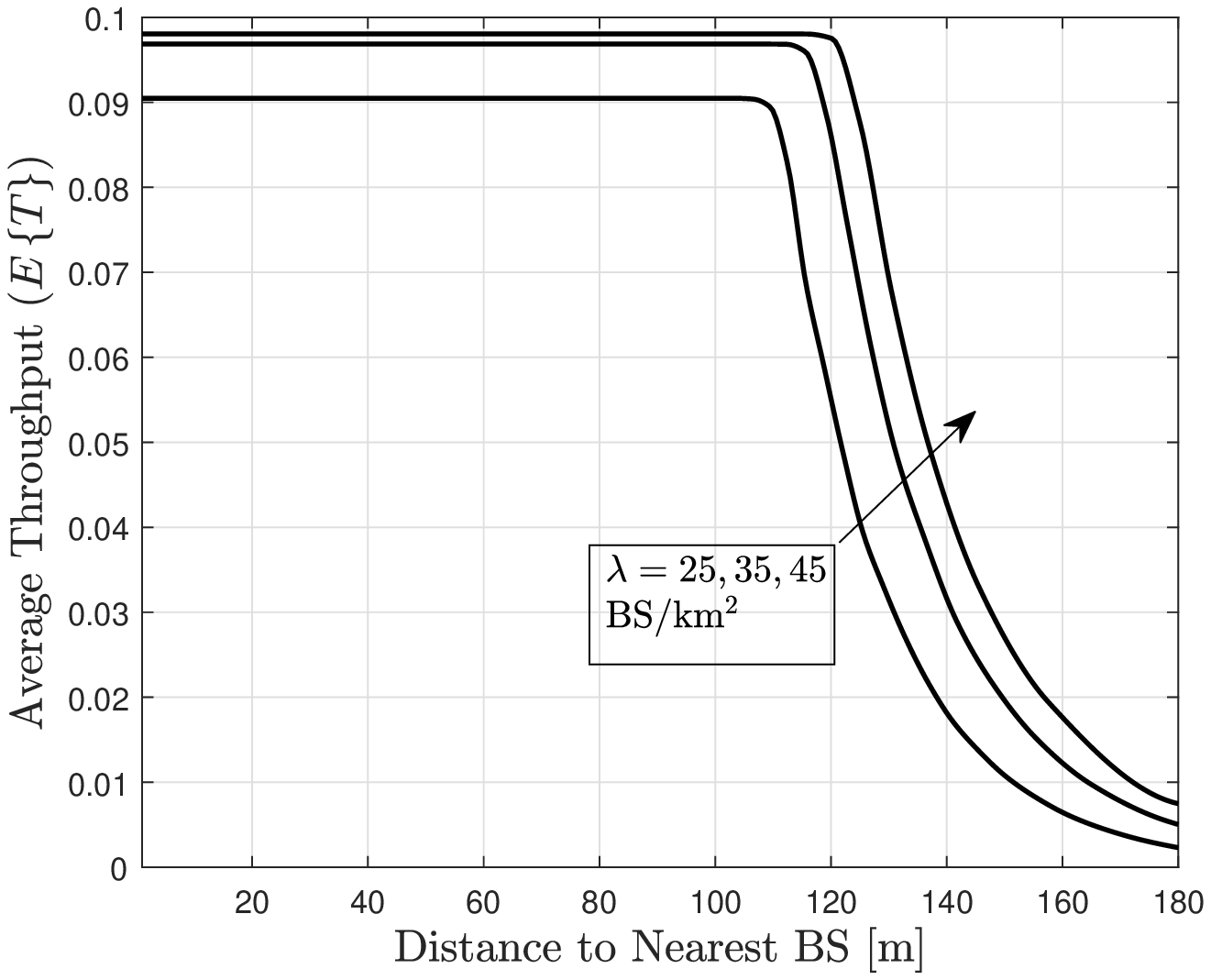}
		\captionof{figure}{ $\;\;$The steady-state average packet throughput as a function of the distance to the nearest BS in meters.}
		\label{fig_thr}
	\end{minipage}
\end{figure}

%

\subsection{The Effect of BSs Densification.} \label{sic:numericalResults1}
\begin{figure}[t!]
	\centering
	\begin{minipage}{.45\textwidth}
		\centering
		\includegraphics[width=2.9 in ]{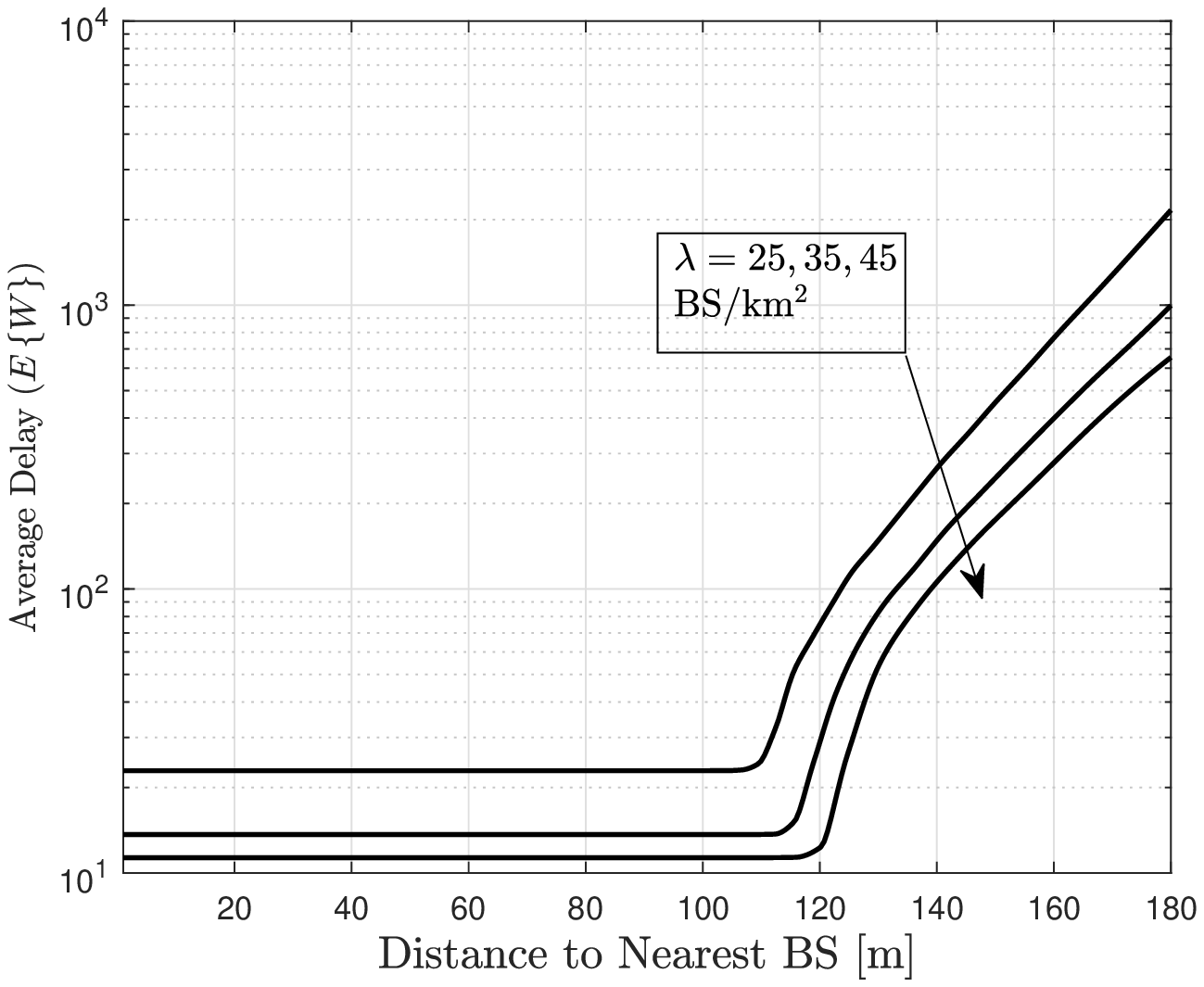}
		\captionof{figure}{$\;\;$The steady-state average delay as a function of the distance to the nearest BS in meters.}
		\label{fig_waiting}
	\end{minipage}%
	\hspace{10mm}
	\begin{minipage}{.45\textwidth}
		\centering
		\includegraphics[width=2.9 in ]{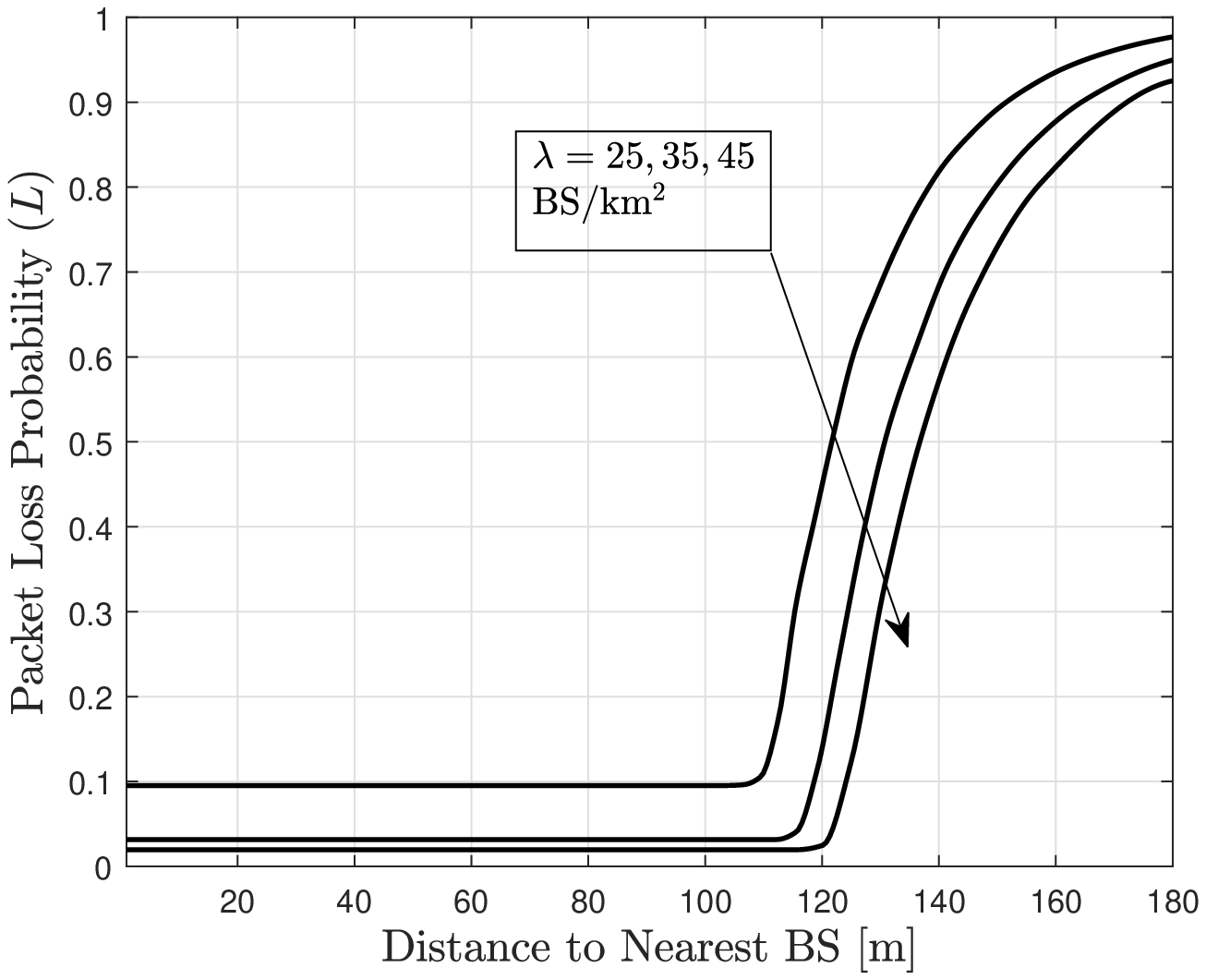}
		\captionof{figure}{ $\;\;$The steady-state packet loss probability as a function of the distance to the nearest BS in meters.}
		\label{fig_loss}
	\end{minipage}
\end{figure}

Next, we depict the performance metrics with BSs densification. { We choose $\lambda\!=\! 35$ BS/km$^2$, $\mu= 175$ device/km$^2$, orthogonal resource blocks $n_c = 64$,    path-loss exponent $\eta \!=\!4$, noise power $\sigma= -120$ dBm, power control threshold $\rho \!=\!-126$ dBm, transmission probability $\Omega= 0.2$,  geometric arrival parameter $a\!=\!0.1$,  $\theta\!=\! -7$ dB, $N=50$ classes, $T_s = 1$ ms, $\zeta P=28$ dBm, buffer size $M=5$, battery size $B= -10$ {dBm-s}, and energy levels $L=25\times 10^3$.} Fig.~\ref{fig_queue} shows  the steady-state average buffer size while  Fig.~\ref{fig_thr} shows the steady-state average packet throughput. Moreover, Fig.~\ref{fig_waiting} depicts the steady-state average delay while Fig.~\ref{fig_loss} depicts the steady-state packet loss probability. It is observed that existence of a cut-off distance between the device and its serving BS, at which the device transition from an eligible transmission state to an energy harvesting state. Moreover, the figures clearly show the performance degradation as the distance towards the nearest BS increases. This performance degradation is mainly because of two main reasons. First, as the distance from the serving BS increases, the amount of harvested energy decreases as depicted in Fig.~\ref{harvest_figure}. Therefore, devices that are closer to their serving BSs harvest energy at a higher rate due to the fixed transmission powers of the BSs and the dominant contribution of the downlink power of the serving BS to the harvested energy. Second, the devices at a larger distance from the serving BS have a higher energy depletion rate due to the higher path-loss that needs to be inverted via the power control in each transmission attempt. As a result, the devices at a high distance from the serving BS have a lower transmission probability and lower throughput, and hence, a higher average buffer, average delay, and packet loss probability. 

 The benefit in terms of increasing the BS intensity ($\lambda$) is further aided by investigating the performance metrics in Figs.~\ref{fig_queue}-\ref{fig_loss}. The effect of the BS densification to serve a certain device intensity is two-fold. First, the BS densification leads to a smaller {mean distance} in \eqref{r_dis}, and hence, the devices would be able to harvest more energy. Second, the smaller distance to the serving BS results in a lower required transmit power, and hence, a higher transmission probability and throughput. Consequently, the lower transmit power for the devices decreases the interference level and yield to a higher transmission success probability and a higher throughput. The key take-away message from the results is that network operation can be always maintained by network densification.

\begin{figure*}[t!]
	\centering
	\begin{subfigure}[t]{0.45\textwidth}
		\centerline{\includegraphics[width=2.9 in ]{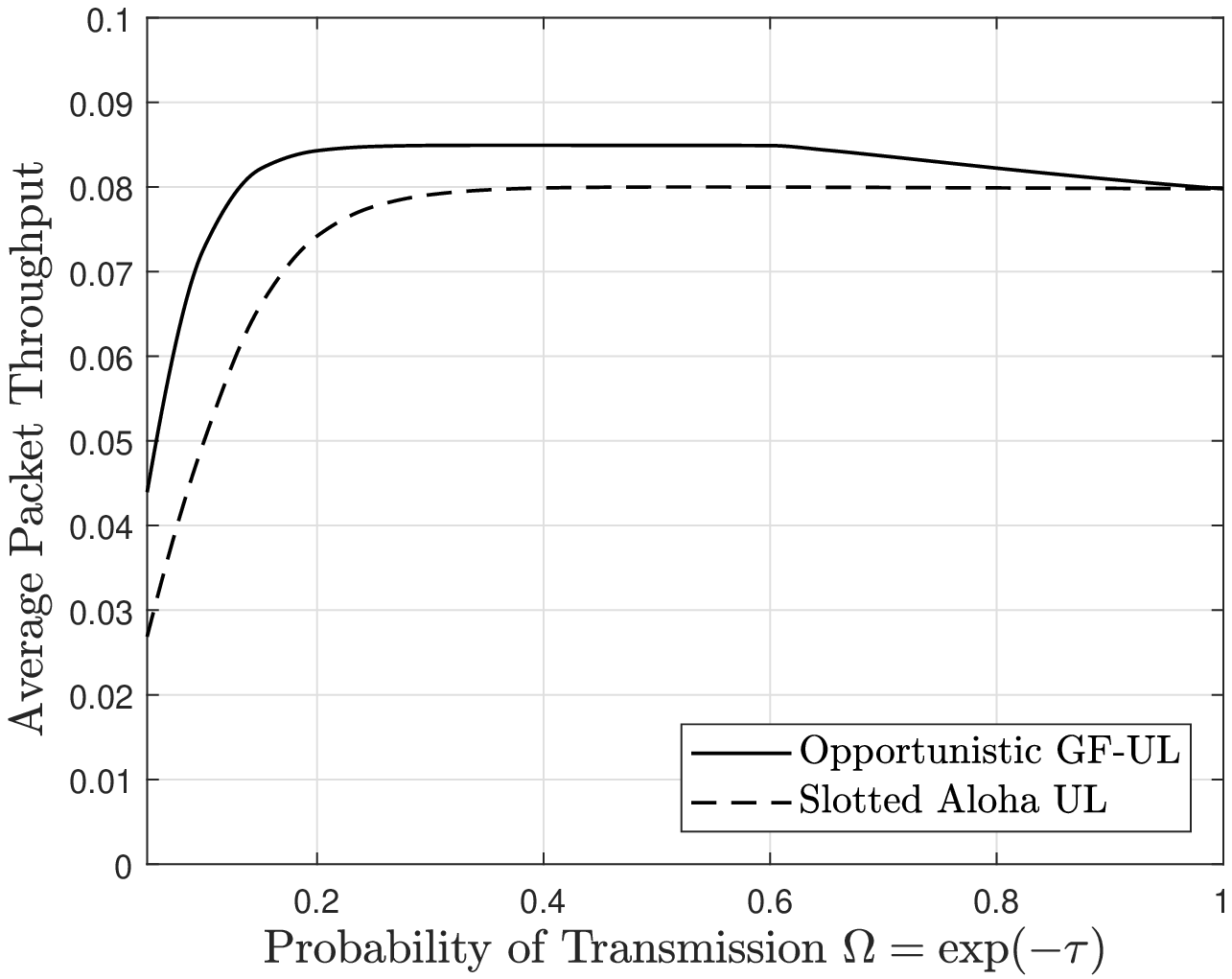}}
		\caption{ $\rho = -124 $dBm.}
	\end{subfigure}
	\vspace{5mm}
	~
	\begin{subfigure}[t]{0.45\textwidth}
		\centerline{\includegraphics[width=2.9 in ]{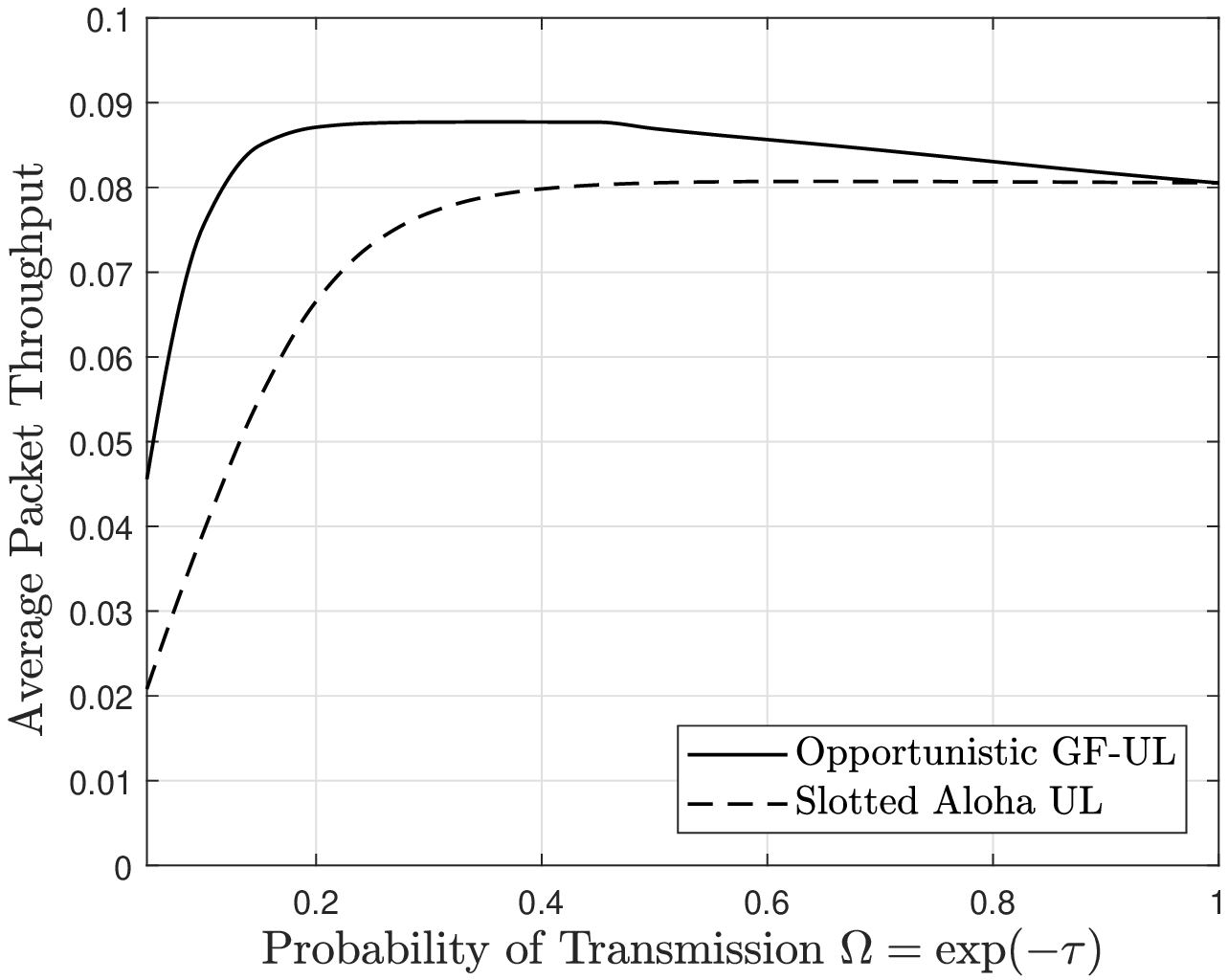}}
		\caption{ $\rho = -126 $dBm.}
	\end{subfigure}
	\begin{subfigure}[t]{0.45\textwidth}
		\centerline{\includegraphics[width=2.9 in ]{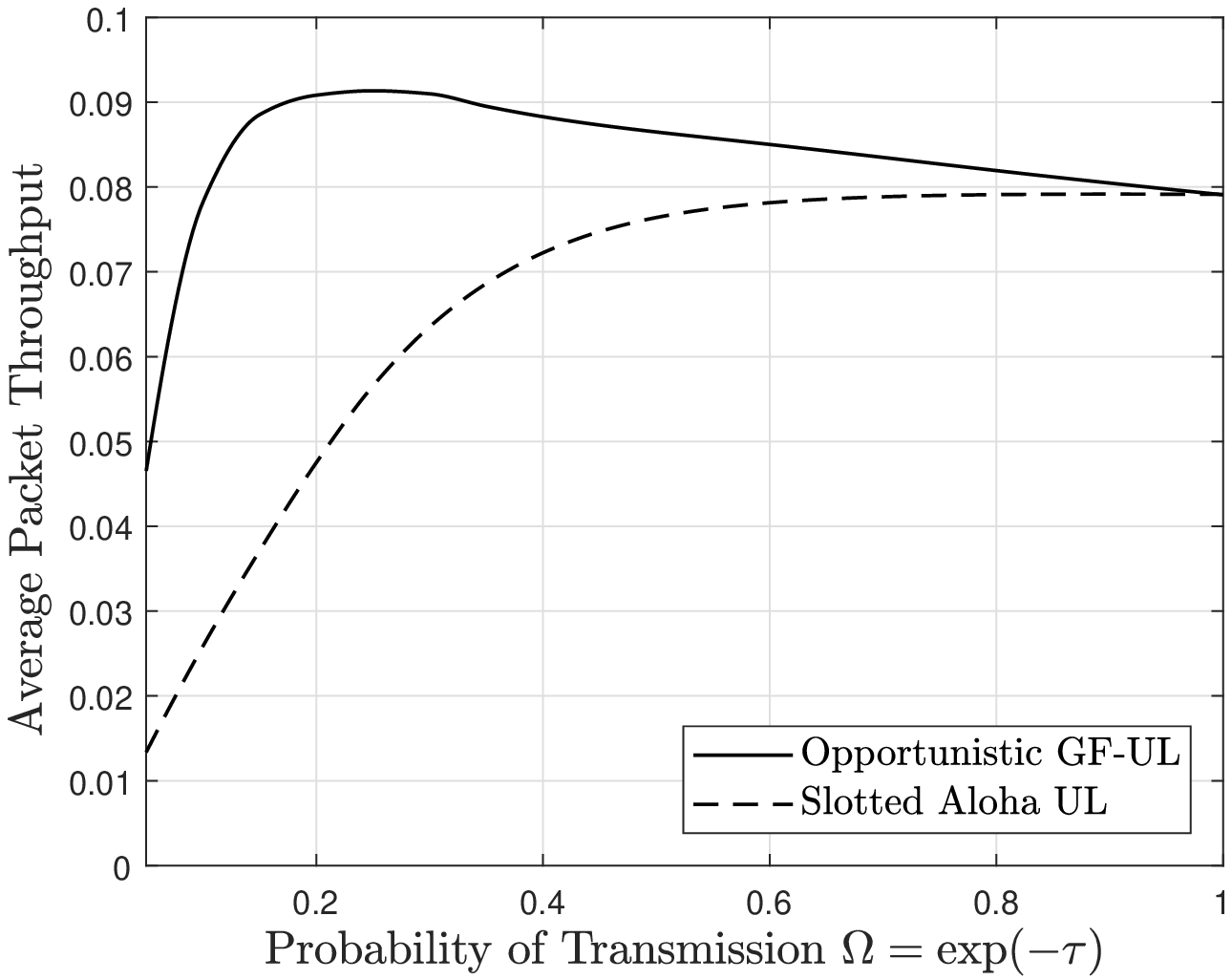}}
		\caption{ $\rho = -128 $dBm}
	\end{subfigure}
	~
	\begin{subfigure}[t]{0.45\textwidth}
		\centerline{\includegraphics[width=2.9 in ]{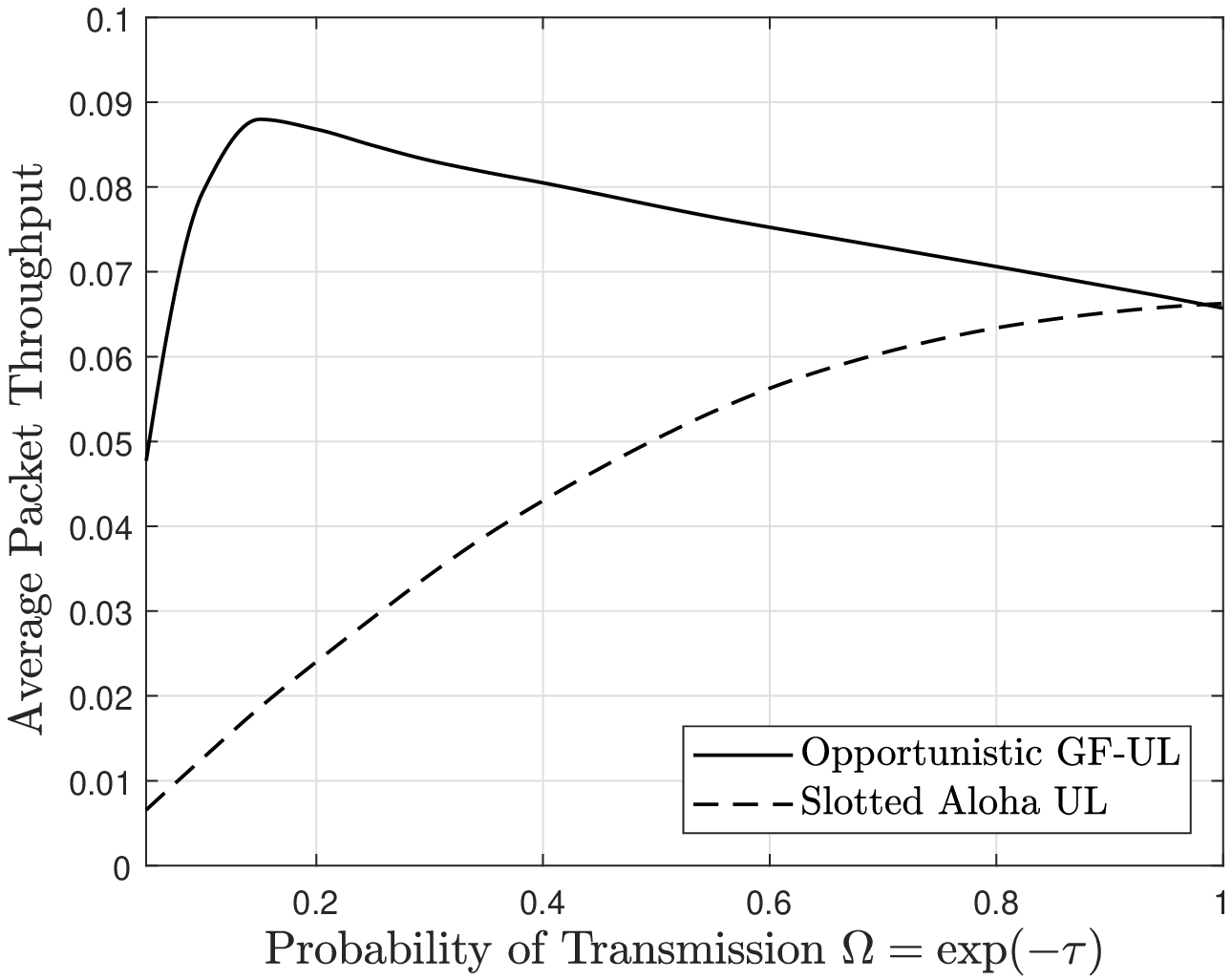}}
		\caption{ $\rho = -130 $dBm.}
	\end{subfigure}
	\caption{ $\;\;$The average packet throughput as a function of the transmission probability ($\Omega$) for various power control thresholds $\rho$ with a slotted Aloha benchmark.}\label{fig_all_rho} 
\end{figure*}

\subsection{Design Insights.} \label{sic:numericalResults2}

{Fig.~\ref{fig_all_rho} depicts the average packet throughput as a function of {the opportunistic probability of transmission $\Omega$}, where the term average here implies averaging the expression in \eqref{ave_throughput} over the $N$ classes. To show-case the gain that can be obtained via opportunistic GF-UL, we are using the equivalent slotted Aloha transmission with the same probability $\Omega$ as a benchmark. The parameters are $\lambda\!=\! 35$ BS/km$^2$,  $\mu= 175$ device/km$^2$, orthogonal resource blocks $n_c = 64$,   path-loss exponent $\eta \!=\!4$,  noise power $\sigma^2= -120$ dBm, power control threshold $\rho \!=\![-124,-130]$ dBm\footnote{It is worth noting that the values of $\rho$ selected are within the practical ranges advertised for IoT applications such as LoRA and Sigfox as stated in \cite{Sundaram2020}. Such high sensitivity is crucial for long range communications with IoT devices with stringent energy constraints.}, geometric arrival parameter $a\!=\!0.1$,  $\theta\!=\! -7$ dB, $N=50$ classes, $T_s = 1$ ms, $\zeta P=28$ dBm, buffer size $M=5$, battery size $B= -10$ {dBm-s}, and energy levels $L=25\times 10^3$. The figures show that the proposed opportunistic GF-UL always outperforms its equivalent slotted Aloha scheme. Moreover, when $\Omega=1$ the two schemes have the same performance because all  $P_T$-capable devices are also Tx-eligible in both schemes.  Looking at the opportunistic GF-UL curves in Fig.~\ref{fig_all_rho}, we can notice that increasing $\rho$ to a certain limit increases the average throughput. In the SINR case, $\rho$ plays a significant role to let the useful transmitted signal suppress the noise, and hence, higher transmission success probability. However, further increasing $\rho$ causes the average throughput to deteriorate. This is due to the fact that the devices will spend more time harvesting the required energy for the transmission without any tangible improvement in the success probability.  Similarly, varying the probability of transmission $\Omega$ provides a trade-off between delay, interference, and efficient utilization of the harvested energy. From the device side, a low value of  $\Omega$ (high value of $\tau$) implies investing the harvested energy in transmissions that are more likely to be successful. From the network perspective, a low value of  $\Omega$ reliefs the aggregate interference by prohibiting transmissions that are more likely to be repeated. However, on the negative side, a low value of  $\Omega$  may also lead to unnecessary transmission deferrals that increase packet delay. Therefore, increasing $\Omega$ to a certain limit increases the average throughput because the opportunistic GF-UL allows only the devices that are more likely to successfully transmit their packet to transmit. Beyond that limit, the opportunistic GF-UL allows more devices that have lower channel gains to contribute to the interference while having unsuccessful transmission attempts. Therefore, for a BSs intensity, there is an optimum pair of ($\rho$ , $\Omega$) that maximizes the average throughput. Hence, that optimum pair minimizes the packet loss probability, the average buffer size, and the average delay. For example, Fig.~\ref{fig_all_rho} shows that setting ($\rho = - 128$ dBm, $\Omega = 0.25$) maximizes the average network throughput.}

\section{Conclusions}\label{sec:Conclusions}
This paper presents a combined stochastic geometry and queuing model for {opportunistic} uplink transmission in wireless-powered IoT networks. The presented model jointly captures temporal traffic generation, the transmission success probability, and the energy harvesting characterization.  A  spatially interacting queues approach is developed, in which the spatial interaction is captured by the mutual interference between the devices.  The results show significant degradation in the performance when the distance between the test device and its serving BS increases. BSs densification is a key solution to improve the network performance. Moreover, the opportunistic GF-UL transmission probability  ($\Omega$)  and the power control parameter ($\rho$) can be jointly optimized to maximize the average packet throughput for a certain BS density. 
\appendix
\subsection{Proof of Lemma 2}\label{proof1}
Starting from rewriting \eqref{SINR_RA} as: 
\begin{align} \label{SINR_RA_rewrite}
p_c=&\mathbb{P}\left\{h_\circ > \frac{\theta}{\rho }(\sigma^2 + \mathcal{I}_{\text{Intra}}+\mathcal{I}_{\text{Inter}}) \;\mid  h_\circ> h_i \;  \forall h_i \in \bold{h}_{\mathcal{V}_\circ} \setminus h_\circ , \forall h_i \in \bold{h}_{\mathcal{V}_\circ}>\tau \right\}\notag \\ 
&\times \mathbb{P}\left\{h_\circ> h_i \;  \forall h_i \in \bold{h}_{\mathcal{V}_\circ} \setminus h_\circ \right\}.
\end{align}
Assuming test cell has a total number of devices equal to  $\mathtt{n}+1$, then the $\mathtt{n}+1$ independent exponentially distributed channels gains have a maximum with the following CCDF:
	\begin{align}\label{eq:AppA_12}
	\bar{F}_{h_{max}} \mid_{\mathcal{N}=\mathtt{n}, h>\tau}\left(h\right)&= 1-(1-\exp\{-(h-\tau)\})^{\mathtt{n}+1}.
	\end{align}
Knowing that $\mathbb{P}\left\{h_\circ> h_i \;  \forall h_i \in \bold{h}_{\mathcal{V}_\circ} \setminus h_\circ \right\}=1/(\mathtt{n}+1)$ and Substituting \eqref{eq:AppA_12} in \eqref{SINR_RA_rewrite}  gives 
		\begin{align}\label{eq:AppA_13}
		p_c \mid_{\mathcal{N}=\mathtt{n}}&= \frac{\mathbb{E}_{\mathcal{I}_{\text{Intera}},\mathcal{I}_{\text{Inter} }}\left\{1-\left( 1-\exp\left\{ -\frac{\theta}{\rho} \left( \sigma^2+\mathcal{I}_{\text{Intra} }+\mathcal{I}_{\text{Inter} } \right) +\tau \right\} \right)^{\mathtt{n}+1}  \right\} }{\mathtt{n}+1}.
		\end{align}
Noting that the PPP is independent in different regions \cite{ martin_book} along with substituting the numerator of \eqref{eq:AppA_13} with its binomial expansion and after applying the total probability theorem  we get \eqref{eq:TX_NB_success}.

{Note that the nearest {\rm BS} association and the employed power control enforce the average inter-cell interference from any interfering device to be less than $\rho$. The aggregated inter-cell interference received at the {\rm BS} from all Tx-eligible and $P_T$-capable devices is obtained as: 
\begin{equation}
\mathcal{I}_{\text{Inter}} =  \sum\limits_{\ell_k \in \bf{\Phi} }  \mathbbm{1}_{\{\mathtt{P}_{k} \left\|u_\circ-\ell_k\right\|^{-\eta} <\rho \}} \mathtt{P}_{k} \mathtt{h}_k \left\|u_\circ-\ell_k\right\|^{-\eta}. \label{eq:inter_proof}	
\end{equation}}

Approximating the set of interfering devices by a PPP with independent transmit powers, the Laplace Transform of \eqref{eq:inter_proof} can be approximated as \eqref{eq:app1}.
\begin{align} \label{eq:app1}
\mathscr{L}_{\mathcal{I}_{\text{Inter}}}(s) \approx \exp \Bigg\{ -2\pi \; \delta\; \Omega \mu^{\prime} \; \mathbb{E}_{\mathcal{P}}\left[\mathcal{P}^{\frac{2}{\eta}} \;   \right] \times \int\limits_{( \rho )^{\frac{-1}{\eta}}}^{\infty}\left(1- \frac{\exp\{-\tau s \; y^{-\eta }\}}{s\;y^{-\eta} +1}\right)\;y\; dy   \Bigg\}, 
\end{align}
\noindent  where, the approximation results form ignoring the correlations between the transmission powers of the devices in the same and adjacent Voronoi cells. The LT is obtained by using Slivnyak's theorem and the probability generating function (PGFL) of the PPP \cite{ martin_book} and following \cite{elsawy2014stochastic}, where the LT is obtained by substituting the value of $\mathbb{E}_{\mathcal{P}}\left[\mathcal{P}^{\frac{2}{\eta}}\right]$ from [Lemma 1,\cite{elsawy2014stochastic}]. To obtain the CDF of the inter-cell interference ( $\mathcal{I}_{\text{Inter}}$), we use Gil-Pelaez theorem \cite{Gill} as
\begin{align}\label{eq:CDF_gil}
\!F_{{\mathcal{I}_{\text{Inter} }}}\!\!\left(x \right)=  &\frac{1}{2}-\frac{1}{\pi} \int\limits_{0}^{\infty} \frac{1}{t} \text{Im} \left\{ \exp\left\{ -j \;t \;x \right\} \mathscr{L}_{\mathcal{I}_{\text{Inter}}}\left(-j\;t\right)\right\} dt.
\end{align}

{For the intra-cell interference, the interference from an interfering device is equal to $\rho$ because of  the nearest {\rm BS} association and the employed power control. For example, the received signal at the test BS  from all $\mathtt{n}+1$ Tx-eligible and $P_T$-capable devices can be expressed by $\{ \mathtt{P}_{\circ} {h}_\circ \mathtt{R}_\circ^{-\eta}, \mathtt{P}_{1} {h}_1 \mathtt{R}_1^{-\eta}, \mathtt{P}_{2} {h}_2 \mathtt{R}_2^{-\eta}, \hdots, \mathtt{P}_{\mathtt{n}} {h}_{\mathtt{n}} \mathtt{R}_{\mathtt{n}}^{-\eta}\}$. Due to the employed power control, the transmit power can be expressed by $\mathtt{P}= \rho \mathtt{R}^{\eta} $. Therefore,  the received signal at the test BS from all the devices can be written as $ \rho \bold{h}_{\mathcal{V}_\circ}=\{\rho {h}_\circ ,\rho h_1, \rho h_2, \hdots, \rho h_{\mathtt{n}}\}$.  As a result, the aggregated intra-cell interference  from $\mathtt{n}$ Tx-eligible and $P_T$-capable interfering devices is obtained as: 
\begin{equation}
\mathcal{I}_{\text{Intra}\mid_{\mathcal{N}=\mathtt{n}}} = \sum\limits_{i \in \mathtt{n} }  \rho \ \mathtt{h}_i. \label{eq:intra_proof}	
\end{equation}}
Let $y$ be the signal that has maximum instantaneous amplitude of those $\mathtt{n}+1$ devices. As such, the LT of each signal other than the maximum is: 
		\begin{align}\label{Laplace_Intra}
		\mathscr{L}_{\mathcal{I}_{\text{\rm Intra}}}\left(s\mid{\mathcal{N}=\mathtt{n}}\right)&= \int_{\tau \rho}^{y} \frac{\frac{1}{\rho} \exp(-\frac{x-\tau\rho}{\rho})}{1-\exp(-\frac{y-\tau\rho}{\rho})} \exp(-sx ) \; dx\notag =\frac{\exp(\frac{y}{\rho}-s \tau\rho )-\exp(-sy+\tau)}{(1+s\rho)(\exp(\frac{y}{\rho})-\exp(\tau))}.
		\end{align}
To characterize the Intra-cell interference conditioned on the devices' number $\mathscr{L}_{\mathcal{I}_{\text{\rm Intra}}}$, \eqref{Laplace_Intra} has to be deconditioned over the RV $y$: 
\begin{align}
\frac{\mathtt{n}+1}{\rho}\exp(-\frac{y-\rho\tau}{\rho})(1-\exp(-\frac{y-\rho\tau}{\rho}))^n,
\end{align}
which yields to, 
		\begin{align}
		\mathscr{L}_{\mathcal{I}_{\text{\rm Intra}}}\left(s\mid{\mathcal{N}=\mathtt{n}}\right)\!&=\left[\exp(\!-s\rho\tau\!)\frac{\mathtt{n}+1}{1+s\rho} \left(\frac{1}{\mathtt{n}} -\frac{\Gamma(\mathtt{n}) \;\Gamma(2+s\rho)}{\Gamma(2+\mathtt{n}+s\rho)} \right)\right]^{\mathtt{n}}.
		\end{align}
Since the summation of exponential random variables follows the gamma distribution, we can approximate the CDF of the aggregated intra-cell interference $F_{{\mathcal{I}_{\rm Intra}}}\!\!\left(x \mid{\mathcal{N}= \mathtt{n}}\right)$ using the moment-matching approach as follows:
\begin{align}
\mathbb{E} \left[ \mathcal{I}_{\rm Intra}\mid_{\mathcal{N}= \mathtt{n}}\right] &= -\lim_{s\to 0} \frac{d \mathscr{L}_{\mathcal{I}_{\text{\rm Intra}}}\left(\!s\mid{\mathcal{N}=\mathtt{n}}\right)}{ds}= \frac{\alpha}{\beta},\label{mn}\\
\mathbb{E} \left[ {\mathcal{I}_{\rm Intra}^2\mid_{\mathcal{N}= \mathtt{n}}}\right] &= \lim_{s\to 0} \frac{d^2 \mathscr{L}_{\mathcal{I}_{\text{\rm Intra}}}\left(\!s\mid{\mathcal{N}=\mathtt{n}}\right)}{{ds}^2}\;\;= \frac{\alpha(1+\alpha)}{\beta^2},\label{vr}
\end{align}
Where $\alpha$ and $\beta$ are, respectively, the shape and the rate parameters for the gamma distribution which has a CDF of $\frac{1}{\Gamma(\alpha)}\gamma(\alpha,\beta x)$. After evaluating the derivatives in \eqref{mn} and \eqref{vr}, $\alpha$ and $\beta$ can be given as in \eqref{eq:shape_par} and \eqref{eq:rate_par}, respectively, which concludes the proof.

\bibliographystyle{IEEEtran}
\balance
\bibliography{refrences,IEEEabrv}
\vfill
\end{document}